\documentclass[prd,aps,twocolumn,a4paper,floatfix]{revtex4-2}

\usepackage{graphicx,psfrag,mathrsfs}
\usepackage{slashed}
\usepackage{mathrsfs}
\usepackage{amsmath,amsfonts,amssymb,amsthm}
\usepackage{hyperref}
\hypersetup{breaklinks=true}
\usepackage{url}
\usepackage{cancel}
\usepackage{accents}
\usepackage{ulem}
\usepackage{xcolor,bbold}

\makeatletter
\newcommand{\sbullet}{%
  \hbox{\fontfamily{lmr}\fontsize{.4\dimexpr(\f@size pt)}{0}
    \selectfont\textbullet}}
\DeclareRobustCommand{\mathbullet}{\accentset{\sbullet}}
\makeatother

\def\p{\partial}

\def\ul{\underline}
\def\nn{\nonumber}

\def\mbn{\mathbullet{\nabla}}

\def\Gb{\mathbullet{\Gamma}}
\def\sD{\slashed{D}}
\def\sg{\slashed{g}}
\def\psib{\ul{\psi}}
\def\sigmab{\ul{\sigma}}
\def\cpr{\mathcal{C}^R_+}
\def\cmr{\mathcal{C}^R_-}
\def\Rc{\mathring{R}}

\newtheorem{remark}{Remark}

\allowdisplaybreaks

\begin{document}

\title{Regularizing Dual-Frame Generalized Harmonic Gauge at Null Infinity}

\author{Miguel Duarte$^{1,2}$}
\author{Justin Feng$^2$}
\author{Edgar Gasper\'in$^{2}$}
\author{David Hilditch$^2$}

\affiliation{$^1$CAMGSD, Departamento de Matem\'atica, Instituto
  Superior T\'ecnico IST, Universidade de Lisboa UL, Avenida Rovisco Pais
  1, 1049 Lisboa, Portugal,\\
  $^2$CENTRA, Departamento de F\'isica, Instituto Superior
  T\'ecnico IST, Universidade de Lisboa UL, Avenida Rovisco Pais 1,
  1049 Lisboa, Portugal,
}

\begin{abstract}
The dual-frame formalism leads to an approach to extend numerical
relativity simulations in generalized harmonic gauge (GHG) all the
way to null infinity. A major setback is that without care, even
simple choices of initial data give rise to logarithmically divergent
terms that would result in irregular variables and equations on the
compactified domain, which would in turn prevent accurate numerical
approximation. It has been shown, however, that a suitable choice of
gauge and constraint addition can be used to prevent their
appearance. Presently we give a first order symmetric hyperbolic
reduction of general relativity in GHG on compactified hyperboloidal
slices that exploits this knowledge and eradicates these log-terms at
leading orders. Because of their effect on the asymptotic solution
space, specific formally singular terms are systematically chosen to
remain. Such formally singular terms have been successfully treated
numerically in toy models and result in a formulation with the
desirable property that unphysical radiation content near infinity is
suppressed.
\end{abstract}

\maketitle

\section{Introduction}

There is strong motivation coming from gravitational wave astronomy to
calculate wave content at future null infinity in asymptotically flat
spacetimes. The relevant geometric notions were pioneered long
ago~\cite{Pen63,NewPen62,BonBurMet62,Sac62,Sac62a}, and so it is no surprise
that there are various proposals to do so. The key idea is always to
compactify-- to draw null infinity to a finite coordinate. Such a
proposal requires picking the character of the domain on which the
equations should be solved. The two possibilities are to use either
outgoing characteristic slices~\cite{Win05} or hyperboloidal slices,
which are spacelike but nevertheless terminate at future null
infinity. For the type of domain chosen, a specific formulation of the
field equations must be specified. The use of compactified coordinates
necessarily means that large quantities enter the game, and need to be
offset against the decay following from asymptotic flatness. Without
care the resulting equations are thus too irregular to be meaningfully
treated either at the PDE or numerical levels. Based on Penrose's idea
of conformal compactification, H\"ubner~\cite{Hub99,Hub01} and
Frauendiener~\cite{DouFra16} make use of the conformal Einstein field
equations (CEFEs), which are explicitly regular, to try to solve this
problem. Although this has shown remarkable
results~\cite{Fri81,Fri81a}, we have not yet figured out how to apply
it to certain spacetimes of interest and in particular, to compact
binary inspiral and merger. Albeit with different applications in
mind, a host of different proposals have been considered for the
regularization on hyperboloidal
slices~\cite{BucPfeBar09,BarSarBuc11,MonRin08,RinMon13}, and treating
specifically the spherically symmetric case
in~\cite{Zen07,VanHusHil14,VanHus14,Van15}. A recurring complication
is the appearance of {\it formally} singular terms which need to be
treated by application of L'H\^opital's rule at null infinity.

Yet another proposal, the subject of this paper, is the dual-frame
approach~\cite{Hil15}, which consists of decoupling the coordinates
from the tensor basis and carefully choosing each. This allows us to
write the Einstein field equations (EFEs) in generalized harmonic
gauge (GHG) and then solve them in hyperboloidal
coordinates~\cite{HilHarBug16,GasHil18,GasGauHil19,GauVanHil21}. This
is dependent on imposing a certain decay of the derivatives of the
radial coordinate light speeds at null infinity, and is deeply related
to the \textit{weak-null condition}~\cite{GasHil18,LinRod03}, a
condition on the non-linearities of quasilinear wave equations
expected to be a sufficient for small data global existence.

The dual foliation (DF) formulation of General Relativity (GR),
together with hyperboloidal coordinates can be used to avoid most of
the formally singular terms~\cite{HilHarBug16}. Although it has been
shown that even the simplest choices of initial data give rise to
terms that diverge with~$\log R$ near null infinity, $R$ being a
suitably defined radial coordinate. These logs create significant
problems in numerical evolutions. However, in~\cite{GasGauHil19} the
authors studied a simple system of wave equations dubbed
\textit{good-bad-ugly} whose non-linearities are known to mimic the
ones present in the EFEs, and found that the logarithmically divergent
terms can be explicitly regularized at first order through a
non-linear change of variables.

Later work around the same model and a generalization to curved
spacetimes resulted in a heuristic method to find asymptotic solutions
and, more importantly, to find which terms may contain logs well
beyond first order~\cite{DuaFenGasHil21}. This idea was subsequently
adapted and applied to GR in~\cite{DuaFenGas22} to study
\textit{peeling}, a requirement on the decay of the components of the
Weyl tensor near null infinity which is a necessary condition for the
smoothness of null infinity~\cite{NewPen62}. The authors found that a
na\"ive choice of gauge and constraint addition prevents the Weyl
tensor from peeling due to the presence of logs. Furthermore, it was
shown that there is special interplay between gauge and constraint
addition which can be exploited to turn all of the metric components
into non-radiating fields at null infinity, except the two that
correspond to gravitational radiation. This idea can be used to
prevent the existence of logs up to arbitrarily high order,
effectively regularizing the EFEs up to that order.

Armed with this knowledge, in this work we build a formulation of GR
in GHG on compactified hyperboloidal slices that provides a set of
regularized equations (in a sense to be clarified) that can be
implemented numerically. We start in section~\ref{section:setup} by
outlining our setup and notation. In
section~\ref{section:gauge_constraint_addition}, convenient gauge
source functions and constraint additions are chosen for each of
the~$10$ metric components. This involves introducing a \textit{gauge
  driver function} that satisfies a wave equation chosen so that it
asymptotes in a certain way. This is a subtle but crucial step to
eradicate the divergent terms. For this reason we end up with~$11$
second order differential equations. In
section~\ref{section:EFEsSecondOrder} we write these equations in a
concise way that is ideal to perform the first order reduction, which
is itself performed in section~\ref{section:Reduction}. By considering
first derivatives of our fields to be evolved variables in their own
right, we find a system of~$55$ regular first order equations whose
principal part can be written in a very compact way. In
section~\ref{section:compactification} we choose a radially
compactified coordinate system and hyperboloidal time coordinate, and
show how that alters the directional derivatives in the system of
equations. Finally on section~\ref{section:EFEsFirstOrder} we present
the full system of first order differential equations and discuss the
existence of formally singular terms, ways to make most of them not
appear in the final equations and the way to deal with the ones that
do. The final system has only two types of formally singular terms,
both of which can be rendered harmless in the numerics, through the
use of L'H\^opital's rule. This amounts to a first order log-free
formulation of the EFEs in GHG and we expect it to serve as an
alternative to the conformal field equations for the inclusion of null
infinity in the computational domain. We wrap up in
section~\ref{section:Hyperbolicity} with a proof that the final system
is symmetric hyperbolic. Concluding remarks are collected in
section~\ref{section:conclusions}.

\section{Framework}\label{section:setup}

To have a self-contained discussion, in this section the general
geometric set up to be used is described in broad terms. Nonetheless,
for a detailed exposition of the geometric framework to be used we
refer the reader to~\cite{DuaFenGasHil21, DuaFenGas22}. Latin indices
will be used as abstract tensor indices while Greek indices will be
used to denote spacetime coordinate indices. $(\mathcal{M},g_{ab})$
will denote a 4-dimensional manifold equipped with a Lorentzian
metric. On~$(\mathcal{M},g_{ab})$ the coordinate
system~$X^{\ul{\alpha}}=(T,X^{\ul{i}})$ will be asymptotically
Cartesian. Let~$\tilde{\nabla}$ denote a flat (and torsionless)
covariant derivative with the key property
that~$\tilde{\nabla}_a\p_{\ul{\alpha}}^b=0$ where $\p_{\ul{\alpha}}$
is the coordinate basis associated to~$X^{\ul{\alpha}}$ (we use the
obvious analogous notation with alternative coordinates). The notion
of flat covariant derivative is simply a way to write partial
derivatives using abstract index notation as it is used in {\tt{xAct}}
---see~\cite{xAct_web_aastex}. Since different coordinate systems will
be used, we make formally identical definitions for the other
coordinate systems.  The coordinates given
by~$X^{\ul{\alpha}'}=(T',X^{\ul{i}'})=(T,R,\theta^A)$, where $R$ is a
radial coordinate defined
via~$R^2=(X^{\ul{1}})^2+(X^{\ul{2}})^2+(X^{\ul{3}})^2$ will be called
Shell coordinates.  Shell coordinates are a generalization of the
standard spherical polar coordinates ---see for
instance~\cite{HilHarBug16}, however, for some of the calculations
discussed in this paper we have used the standard coordinatization
of~$\mathbb{S}^2$ so that the Shell coordinates are simply spherical
polar.

Since the difference between two connections is a tensor,
the Christoffel transition tensor of a given pair of covariant
derivatives is fixed by the relation,
\begin{align}\label{eqn:Gammadef}
  \Gamma[\nabla,\mathbullet{\nabla}]_{a}{}^{b}{}_{c}v^c =
  \nabla_av^b - \mbn_av^b\,,
\end{align}
where~$v^a$ is a vector. The transition
tensor~$\Gamma[\nabla,\mbn]_{a}{}^{b}{}_{c}$ is defined analogously.

We will use~\eqref{eqn:Gammadef} extensively hence to have a simpler
notation we define
\begin{align}
  \Gb_{a}{}^{b}{}_{c} :=
  \Gamma[\nabla,\mathbullet{\nabla}]_{a}{}^{b}{}_{c}\,.
\end{align}

\subsection{Representation of the metric}

Introduce the following vectors
\begin{align}\label{eqn:psiinshellchart}
  \psi^a&=\p_T^a+\mathcal{C}_+^R\p_R^a\,,\quad
  \ul{\psi}^a=\p_T^a+\mathcal{C}_-^R\p_R^a\,,
\end{align}
where~$\mathcal{C}_+^R$ and~$\mathcal{C}_-^R$ are fixed by the
requirement that~$\psi^a$ and~$\ul{\psi}^a$ are null vectors with
respect to the metric~$g_{ab}$. With these conventions~$\psi^a$
and~$\bar{\psi}^a$ correspond to outgoing and incoming null vectors
respectively.

Additionally let
\begin{align}\label{eqn:xitoeta}
  \sigma_a&=e^{-\varphi}\psi_a\,,\quad
  \ul{\sigma}_a=e^{-\varphi}\ul{\psi}_a\,,
\end{align}
where~$\varphi$ is fixed by requiring that
\begin{align}
  \sigma_a\p_R^a=-\ul{\sigma}_a\p_R^a=1\,,
\end{align}
so we can write,
\begin{align}\label{eqn:xiinshellchart}
  \sigma_a&=-\mathcal{C}_+^R\nabla_aT+\nabla_aR
  + \mathcal{C}^+_A\nabla_a\theta^A\,,\nn\\
  \ul{\sigma}_a&=\mathcal{C}_-^R\nabla_aT-\nabla_aR
  + \mathcal{C}^-_A\nabla_a\theta^A\,.
\end{align}
With these elements the inverse metric is written as
\begin{align}\label{eqn:metricrepresentation}
g^{ab}=-2\tau^{-1}e^{-\varphi}\,\psi^{(a}\ul{\psi}^{b)}+\slashed{g}^{ab} \,,
\end{align}
where the null vectors satisfy,
\begin{align}
  &\sigma_a\psi^a=\ul{\sigma}_a\ul{\psi}^a=0\,,\quad
  \sigma_a\ul{\psi}^a=\ul{\sigma}_a\psi^a=-\tau\,,
\end{align}
where we define~$\tau:=\mathcal{C}_+^R-\mathcal{C}_-^R$.
In~\eqref{eqn:metricrepresentation}, the normalization ensures that
\begin{align}
  \slashed{g}^{ab}\sigma_b=\slashed{g}^{ab}\ul{\sigma}_b=0\,,
\end{align}
where ~$\slashed{g}_a{}^b$ acts as a projection operator orthogonal to
these two covectors. Note that~$\slashed{g}^{ab}$ is not the inverse
induced metric on the surfaces of constant~$T$ and~$R$, as it is not
orthogonal to~$\nabla_aT$ or~$\nabla_aR$, but rather to~$\sigma_a$
and~$\ul{\sigma}_a$. The covariant version of metric can be written
naturally as,
\begin{align}
  g_{ab}=-2\tau^{-1}e^{\varphi}\,\sigma_{(a}\ul{\sigma}_{b)}+\slashed{g}_{ab}\,.
\end{align}
Moreover, for the sector ~$\sg^{ab}$, we introduce the following split,
\begin{align}\label{eqn:qconformaltransformation}
  q_{ab}= \mathring{R}^{-2}\slashed{g}_{ab}
  \,, \quad (q^{-1})^{ab} = \mathring{R}^{2}\slashed{g}^{ab}\,,
\end{align}
with 
\begin{align}
  \mathring{R}^2 := R^2 e^\epsilon :=
  R^2\sqrt{|\slashed{g}|/|\slashed{\mathbullet{g}}|}
\end{align}
where~$\Rc$ is known as the areal radius
and~$|\slashed{\mathbullet{g}}|$ represents the determinant of the
standard metric of~$\mathbb{S}^2$ of radius~$R$.
In~\cite{DuaFenGas22} the areal radius is never used, however, it
plays a central role in the calculations of this paper. The use
of~$\Rc$ can be thought then as a further refinement of the geometric
framework of~\cite{DuaFenGas22}.  Additionally, to simplify the
notation we also define~$\mathbullet{\epsilon} := \tfrac{1}{2}\ln
|\slashed{\mathbullet{g}}|$.  Observe that the determinant of~$q$ is
that of the metric of the unit~$\mathbb{S}^2$ in shell
coordinates. Finally, we use a particular parameterization of the
angular part of~$(q^{-1})^{ab}$, which in shell coordinates reads
\begin{align}
  (q^{-1})^{AB} = \begin{bmatrix}
    e^{-h_+}\cosh h_\times & \frac{\sinh h_\times}{\sin \theta} \\
    \frac{\sinh h_\times}{\sin \theta}
    & \frac{e^{h_+}\cosh h_\times}{\sin^2 \theta}
  \end{bmatrix}\,.
\end{align}
Despite that no linearization is intended, in the last equation we use
the symbols~$h_\times$ and~$h_+$ to denote the two degrees of freedom
of gravitational waves.

In this work we will also use the components of~$\sg_a{}^b$ in mixed
form, so we write the non-zero ones here,
\begin{align}\label{sgMixed}
  \sg_A{}^T = \frac{\mathcal{C}_A}{\tau}\,,\quad\sg_A{}^R
  = \frac{\mathcal{C}^+_A\cpr + \mathcal{C}^-_A\cmr }{\tau}
  \,,\quad\sg_A{}^B=\delta_A^B\,.
\end{align}
The metric is thus represented by,
\begin{align}\label{eqn:BasicMetricVariables}
   \mathcal{C}_\pm^R\,, \quad \mathcal{C}^\pm_A\,, \quad
   \varphi\,, \quad \mathring{R}\,, \quad h_+\,, \quad h_\times\,.
\end{align}

\paragraph*{The projected covariant derivative:}
We define the derivative~$\sD_A$ for scalar functions $\phi$,
\begin{align}\label{eqn:sDDef}
\sD_A \phi:= \slashed{g}_{A}{}^{b}\mbn_b \phi\,.
\end{align}
We write the inverse conformal metric as~$ (q^{-1})^{ab} $ to stress
that, in general, to raise and lower indices we do not use $q^{ab}$
but $\sg^{ab}$ instead. Namely, $q^{ab} =
\slashed{g}^{ac}\slashed{g}^{bd}q_{cd} \neq (q^{-1})^{ab}$. However,
for simplicity of the expressions, the index associated with this
derivative and only that index will be raised with~$\sg^{ab}$. We
wrote the lower index in~\eqref{eqn:sDDef} as an angular one
since~$\slashed{g}_{a}{}^{b}\p_T^a \neq 0$
and~$\slashed{g}_{a}{}^{b}\p_R^a \neq 0$. In fact~$\sD_A\phi$ can be
written as,
\begin{align}
  \sD_A\phi = \frac{1}{\tau}(\mathcal{C}_A^+\nabla_\psi\phi
  +\mathcal{C}_A^-\nabla_{\psib}\phi)+\p_{\theta^A}\phi\,.
\end{align}
We define the vector field~$T^a:=\p_T^a$ and denote the covariant
derivative in the direction of~$T^a$ as~$\nabla_T$. Analogous notation
will be used for directional derivatives along other vector
fields. Since in expressions to be derived, terms of the
form~$R^{-n}(\log R)^m$ will appear, we make clear at this point that
the phrase 'order $n$' will mean terms proportional to~$R^{-n}$.

\paragraph*{Representation of the connection:}
A calculation using
\begin{align}
  \Gb_a{}^b{}_c = \frac{1}{2}\sg^{bd}
  (\mbn_a \sg_{db}+\mbn_b\sg_{ad}-\mbn_d \sg_{ab})\,,
\end{align}
gives
\begin{align}\label{eqn:ConnectionComps}
	&\nabla_a \mathcal{C}_+^R =
  -\Gb_a{}^\sigma{}_\psi\,,\quad\nabla_a \mathcal{C}_-^R =
  \Gb_a{}^{\sigmab}{}_{\psib}\,,\nn\\
  &\nabla_a \mathcal{C}^+_A = 2\sg_{bA}\Gb_a{}^{(b
    \sigma)}
  -\frac{\mathcal{C}^+_A+\mathcal{C}^-_A}{\tau}\Gb_a{}^\sigma{}_\psi\,,
  \nn\\
  &\nabla_a\mathcal{C}^-_A = 2\sg_{bA}\Gb_a{}^{(b
    \sigmab)}-\frac{\mathcal{C}^+_A+\mathcal{C}^-_A}{\tau}
  \Gb_a{}^{\sigmab}{}_{\psib}\nn\,,\\
  &\nabla_a\varphi = \Gb_a{}^b{}_c(\delta_b^c-\sg_b{}^c)\,,\nabla_a \sg^{AB} =
  -2\sg_b{}^A\sg_c{}^B\Gb^{(bc)}{}_a\,,\nn\\
  &\nabla_a (\epsilon + \mathbullet{\epsilon}) =\sg_a{}^b\Gb_b{}^a{}_c\,.
\end{align}
These expressions will serve as shorthands to write the EFE in a more
compact way. Similarly one can write the derivatives of metric
components in terms of~$\Gb_a{}^b{}_c$. For brevity these expressions
are omitted. Note that these covariant derivatives are interchangeable
with~$\mbn$ even when they act upon~$\mathcal{C}^\pm_A$, as~$A$ is not
a tensorial index but rather a coordinate one.

\subsection{The good, the bad, the ugly and the stratified null forms}

The term \textit{stratified null forms} (SNFs) will refer to
expressions involving products of terms containing at most one
derivative of the evolved fields and having a faster decay
than~$R^{-2}$ close to $\mathscr{I}^{+}$.  As was shown
in~\cite{DuaFenGasHil21} and~\cite{DuaFenGas22}, this definition is
important because it includes every term that cannot possibly
interfere with the first order asymptotics of a good-bad-ugly system,
which we introduce below. This allows us to categorize terms in an
effective way according to their influence on the leading order
behavior at null infinity. Throughout this work we will use a
calligraphic~$\mathcal{N}_\phi$ to denote stratified null forms,
where~$\phi$ is the field whose evolution equation
contains~$\mathcal{N}_\phi$. We introduce the good-bad-ugly model,
\begin{align}
  & \square g = \mathcal{N}_g\,,\nn\\
  &\square b = (\nabla_T g)^2 + \mathcal{N}_b\,,\nn\\
  & \square u = \tfrac{2}{R}\nabla_T u + \mathcal{N}_u\,,
  \label{eqn:gbu}
\end{align}
Here, $\square$ is defined as~$g^{ab}\nabla_a\nabla_b$ and an
analogous definition holds for~$\mathbullet{\square}$. Good fields are
characterized asymptotically by a leading order term with no logs
whose decay improves under derivatives along outgoing null curves and
does not under derivatives along incoming ones. Bad fields have a
leading term proportional to~$\log R$ and behave similarly under null
derivatives. Ugly fields in general have logs in subleading terms and
their decay improves under derivatives of any kind. In its simplest
form, this system has been studied extensively as a toy model for the
EFEs because the latter can be written in such a way that its leading
order non-linearities are mimicked by those present on the RHSs
of~\eqref{eqn:gbu}. In fact if we consider wave operators in curved
spacetimes and allow the existence of more than one good and one ugly,
The EFEs can be written in exactly this form. Even in flat spacetimes
and with~$\mathcal{N}_\phi=0$, this system is known to give rise to
logarithmically divergent terms at null infinity. However it is shown
in~\cite{DuaFenGas22} that gauge picking and constraint addition can
be used to prevent the appearance of those logs up to an arbitrarily
high order. This is what we will explore throughout this work.

\paragraph*{Ugly equation with p:} It is convenient to introduce
different wave operators that will make expressions more tractable
throughout this work. We define the shell wave operator as,
\begin{align}
  \mathbullet{\square} \phi =
  g^{ab}\mathbullet{\nabla}_a\nabla_b\phi\,.
\end{align}
The ugly equation with a natural number~$p$, as defined
in~\cite{DuaFenGas22}, can be written in the form,
\begin{align}\label{eqn:superugly}
  \mathbullet{\square}u = \frac{2(p+1)}{R}\nabla_Tu + \mathcal{N}_u\,,
\end{align}
but it can also be written in an equivalent, yet more convenient way,
\begin{align}\label{eqn:superugly2}
  \mathbullet{\square}u = \frac{2(p+1)e^{-\varphi}}{\tau\Rc}
  \nabla_\psi\Rc\nabla_{\psib}u + \mathcal{N}_u\,,
\end{align}
by simply redefining~$\mathcal{N}_u$ The shell wave operator can be
expanded as,
\begin{align}\label{eqn:lhs2}
\mathbullet{\square} u = &-
\frac{2e^{-\varphi}}{\tau}\nabla_{\psi}\nabla_{\psib} u
+ \frac{2e^{-\varphi}}{\tau}(\mbn_\psi\psib)^a\nabla_au
+ \sg^{ab}\mbn_a\nabla_bu\nn\\
=&-\frac{2e^{-\varphi}}{\tau}\nabla_{\psi}\nabla_{\psib} u
+ \frac{2e^{-\varphi}}{\tau^2}
\nabla_\psi\cmr(\nabla_\psi u-\nabla_{\psib}u)
+\mathbullet{\cancel{\square}}u
\end{align}
where~$\mathbullet{\cancel{\square}}u := \sg^{ab}\mbn_a\nabla_bu$. For
conciseness we define the second order differential
operator~$\mathbullet{\square}_p$ as,
\begin{align}\label{boxp}
  \mathbullet{\square}_pu &:=
  \mathbullet{\square}u - \frac{2(p+1)e^{-\varphi}}{\tau\Rc}
  \nabla_\psi\Rc\nabla_{\psib}u\nn\\
  &\,= -\frac{2e^{-\varphi}}{\tau \Rc^{p+1}}\nabla_\psi
  \left[\Rc^{p+1}\nabla_{\psib}u\right]
  + \mathbullet{\cancel{\square}}u \nn\\
  &\,+ \frac{2e^{-\varphi}}{\tau^2}\nabla_\psi\cmr
  (\nabla_\psi u-\nabla_{\psib}u)\,.
\end{align}
Putting~\eqref{eqn:superugly} and~\eqref{eqn:lhs2} together we get
that every ugly equation with a natural number~$p$ can be written in
the form,
\begin{align}\label{uEqShort}
  \mathbullet{\square}_pu = \mathcal{N}_u\,.
\end{align}

\paragraph*{Good equation:} The general form of a good equation
can be obtained straightforwardly from the ugly
one~\eqref{eqn:superugly} by setting~$p=0$. Therefore if~$g$ is good,
it satisfies the following equation,
\begin{align}
  -\frac{2e^{-\varphi}}{\tau\Rc}\nabla_\psi&\left[\Rc\nabla_{\psib}g\right]
  + \mathbullet{\cancel{\square}}g =\\
  &-\frac{2e^{-\varphi}}{\tau^2}\nabla_\psi\cmr(\nabla_\psi g-\nabla_{\psib}g)
  + \mathcal{N}_g\nn\,,
\end{align}
or more simply,
\begin{align}\label{gEqShort}
  \mathbullet{\square}_0 g = \mathcal{N}_g\,.
\end{align}

\paragraph*{Alternative form:} Another way to write the equations
is to push as many SNFs as possible to the RHS and keep on the LHS
only the terms which are either principal or contribute to leading
order. For this we need to rewrite~$\mathbullet{\cancel{\square}}$,
\begin{align}
  \mathbullet{\cancel{\square}}u &=
  \frac{1}{\tau}\left(\frac{\mathcal{C}_A}{\tau}\sD^A\cpr
  -\sD^A\mathcal{C}^+_A\right)\nabla_{\psib}u \\
  &- \frac{1}{\tau}\left(\frac{\mathcal{C}_A}{\tau}\sD^A\cmr
  +\sD^A\mathcal{C}^-_A\right)\nabla_{\psi}u +\slashed{\Delta}u\,,\nn
\end{align}
where~$\mathcal{C}_A:=\mathcal{C}_A^++\mathcal{C}_A^-$
and~$\slashed{\Delta}u:=\sD^A\sD_Au$. All terms on the RHS are SNFs
except the last one, so we write~\eqref{uEqShort} as,
\begin{align}\label{uEqLong}
  \frac{1}{\Rc^{p+1}}\nabla_\psi\left[\Rc^{p+1}\nabla_{\psib}u\right]
  - \frac{\tau e^\varphi}{2}\slashed{\Delta}u =\tilde{\mathcal{N}}_u
\end{align}
with the RHS being a set of null forms,
\begin{align}
  \tilde{\mathcal{N}}_u = &
  -\frac{\tau e^\varphi}{2}\mathcal{N}_u
  +\frac{e^\varphi}{2}\left(\frac{\mathcal{C}_A}{\tau}
  \sD^A\cpr-\sD^A\mathcal{C}^+_A\right)\nabla_{\psib}u\nn \\
  &- \frac{e^\varphi}{2}\left(\frac{\mathcal{C}_A}{\tau}
  \sD^A\cmr+\sD^A\mathcal{C}^-_A\right)\nabla_{\psi}u\nn\\
  &+ \frac{1}{\tau}\nabla_\psi\cmr(\nabla_\psi u-\nabla_{\psib}u)\,.
\end{align}
Naturally, the version of~\eqref{uEqLong} for the good fields is
obtained simply by setting~$p=0$.

\subsection{Asymptotic flatness at null infinity}

Astrophysically relevant objects are modeled in General Relativity via
asymptotically flat spacetimes. Although, from a physical perspective
the notion of asymptotic flatness is clear: it represents the
requirement that the metric asymptotes to the Minkowski spacetime in
far regions of the spacetime, from a mathematical perspective there
are several related but not necessarily equivalent definitions. Our
working notion of asymptotic flatness near null infinity is that,
\begin{align}
  \gamma|_{\mathscr{I}^+}=0\,,
\end{align}
where we define
\begin{align}
  &\varphi = \gamma_1\,,\quad
  \mathcal{C}_\pm^R = \pm1
  + \gamma^\pm_2\,,\quad
  \mathcal{C}_A^\pm = \Rc\gamma_3^\pm\,, \nn \\
  &\Rc^{-1} = \gamma_4\,,\quad
  h_+ = \gamma_5\,,\quad
  h_\times = \gamma_6\,.
\end{align}

\subsection{Initial data}

Let~$\mathcal{S}$ denote the Cauchy surface defined by~$T=T_0$,
with~$T_0$ constant. In~\cite{DuaFenGas22}, we considered initial data
with the following fall-off at spatial infinity, 
\begin{align}\label{assumptionTimeder}
  &\gamma\rvert_{\mathcal{S}}=
  \sum_{n=1}^{\infty}\frac{m_{\gamma,n}}{\Rc^n}\,,\quad
  \nabla_Tg\rvert_{\mathcal{S}}=
  O_{\mathcal{S}}(\Rc^{-2})\,,
\end{align}
where~$m_{\gamma,n}$ denote functions that only depend on the angular
coordinates $\theta^A$. Here the subscript~$\mathcal{S}$ is placed to
stress that this decay is assumed only on the initial hypersurface and
not close to null infinity. Although this data is not the most general
one, this class of initial data is large enough to admit non-vanishing
ADM mass and linear momentum. Moreover, it includes initial data of
physically relevant spacetimes. In practice for the results in this
paper this assumption could be relaxed to,
\begin{align}
  &\gamma\rvert_{\mathcal{S}}=
  \frac{m_{\gamma,1}}{\Rc}+O_{\mathcal{S}}(\Rc^{-2})\,,\quad
  \nabla_Tg\rvert_{\mathcal{S}}=
  O_{\mathcal{S}}(\Rc^{-2})\,,
\end{align}
without further change.

\paragraph*{Permissible coordinate changes:} There is not one universal
definition of asymptotic flatness at spatial infinity, rather there is
an interplay between the field equations, the physics under
consideration, and the rate at which the metric becomes flat. A weak
definition thereof is,
\begin{align}\label{asympflatspatial}
  g_{\ul{\alpha \beta}}=\eta_{\ul{\alpha \beta}}+O_p(R^{-1})\,,
\end{align}
where~$\eta_{\ul{\alpha \beta}}$ is the Minkowski metric, $p\geq 1$
and~$O_p(R^{-m})$ means that its partial
derivatives~$\p_{\ul{\alpha}}$ of order~$n$ decay as~$R^{-n-m}$ for
all~$n=1,\dots,p$. Note that our
requirements~\eqref{assumptionTimeder} satisfy this definition of
asymptotic flatness. It has been shown in~\cite{DuaHil19} that a
coordinate transformation is \textit{permissible}, meaning it
preserves~\eqref{asympflatspatial}, if and only if it is
asymptotically a Poincar\'e transformation in the following sense,
\begin{align}\label{permissible}
  X^{\ul{\alpha}}=\Lambda^{\ul{\alpha}}{}_\alpha x^{\alpha}
  + c^{\ul{\alpha}}(\theta,\phi)+O_{p+1}(\tilde{r}^{-1})\,,
\end{align}
where~$\Lambda^{\ul{\alpha}}{}_\alpha$ is a Lorenz transformation
and~$x^{\alpha}$ is another asymptotically Cartesian coordinate system
with a radial coordinate~$\tilde{r}$ built in the usual way. It is
interesting to see how a permissible coordinate change affects our
initial data. In many spacetimes of interest, the
functions~$m_{\gamma,1} = m$, where~$m$ is a constant, e.g the Kerr
metric in Boyer-Lindquist coordinates. If we boost this metric with a
generic permissible coordinate change, the new
functions~$m_{\tilde{\gamma},1}$ will pick up dependencies on the
angles at leading order in~$\Rc^{-1}$, that is, $m_{\tilde{\gamma},1}
= m_{\tilde{\gamma},1}(m,\theta,\phi)$. Note that the initial data
that one gets after such a boost, still satisfies the
requirements~\eqref{asympflatspatial} and hence can be used evolved
numerically with the method we will present throughout this work.

\section{Gauge and constraint addition}
\label{section:gauge_constraint_addition}

In~\cite{DuaFenGas22}, the vacuum EFEs were derived as a set of
non-linear wave equations satisfied by the
variables~\eqref{eqn:BasicMetricVariables}. We present them here as
concisely as possible while still keeping the freedom to choose the
gauge and the constraint addition in each of the equations,
\begin{align}\label{EFE}
  &\mathbullet{\square} \mathcal{C}_+^R =
  e^{-\varphi}\tilde{\mathcal{R}}_{\psi\psi}
  + (\Gamma\Gamma)_{\psi\psi}\nn\,,\\
  &\mathbullet{\square} \mathcal{C}_-^R =
  -e^{-\varphi}\tilde{\mathcal{R}}_{\ul{\psi}\ul{\psi}}
  + (\Gamma\Gamma)_{\ul{\psi}\ul{\psi}}\nn\,,\\
  &\mathbullet{\square} \varphi =
  \frac{2e^{-\varphi}}{\tau}\tilde{\mathcal{R}}_{\psi\psib}
  +(\Gamma\Gamma)_{\psi\ul{\psi}}\nn\,,\\
  &\mathbullet{\square} \mathcal{C}^+_A =
  -e^{-\varphi}\tilde{\mathcal{R}}_{\psi A}
  + \frac{\bar{\mathcal{C}}_A}{\tau}\mathbullet{\square}
  \mathcal{C}_+^R+(\Gamma\Gamma)_{\psi A}\nn\,,\\
  &\mathbullet{\square} \mathcal{C}^-_A =
  -e^{-\varphi}\tilde{\mathcal{R}}_{\psib A}
  - \frac{\bar{\mathcal{C}}_A}{\tau}\mathbullet{\square}\mathcal{C}_-^R
  +(\Gamma\Gamma)_{\ul{\psi} A}\nn\,,\\
  &\mathbullet{\square} (\epsilon + \mathbullet{\epsilon})
  = -\tilde{\slashed{\mathcal{R}}}+(\cancel{\Gamma\Gamma})\,,\nn\\
  &\mathbullet{\square}h_+ = -\frac{2}{\sg^{\theta\theta}}
  \tilde{\mathcal{R}}^{\theta\theta}
  + (\Gamma\Gamma)^{\theta\theta}\nn\,,\\
  &\mathbullet{\square}h_\times =
  -2\sg_{\theta\phi}\coth h_\times\tilde{\mathcal{R}}^{\theta\phi}
  + (\Gamma\Gamma)^{\theta\phi}\,.
\end{align}
where the different components of~$(\Gamma\Gamma)_{ab}$ conceal the
complicated non-linearities,
\begin{align}
  &(\Gamma\Gamma)_{\psi\psi}:=
  \frac{2}{\tau}\Gb_\psi{}^\sigma{}_a\Gb_\psi{}^{\sigma a}
  -e^{-\varphi}\Gb_\psi{}^b{}_a\Gb_{\psi{}b}{}^a
  - 2\sg_a{}^c\Gb_\psi{}^a{}_b\Gb_c{}^{\sigma b}\,,\nn\\
  &(\Gamma\Gamma)_{\psib\psib}:=
  - \frac{2}{\tau}\Gb_{\psib}{}^{\sigmab}{}_a\Gb_{\psib}{}^{{\sigmab} a}
  +e^{-\varphi}\Gb_{\psib}{}^b{}_a\Gb_{\psib{}b}{}^a\nn\\
  &\quad\quad\quad\quad\,\,\,
  + 2\sg_a{}^c\Gb_{\psib}{}^a{}_b\Gb_c{}^{\sigmab b}\nn\,,\\
  &\begin{aligned}
     (\Gamma\Gamma)_{\psi\ul{\psi}}:=&
     - \frac{2}{\tau^2}\left[\Gb_{\psib}{}^a{}_{(\psi}\Gb_{\psib)}{}^{\sigma}{}_a
       + \Gb_{\psi}{}^a{}_{(\psib}\Gb_{\psi)}{}^{\sigmab}{}_a\right]\nn\\
     &+\Gb_a{}^{db}\Gb^{ac}{}_e\left[2\sg_{cb}(\delta_d^e-\sg_d{}^e)
       +\sg_{cd}(\delta_b^e-\sg_b{}^e)\right]\,,\nn\\
   \end{aligned}\\
  &\begin{aligned}
     (\Gamma\Gamma)_{\psi A}:=&
     \frac{2}{\tau}\Gb^\sigma{}_A{}^{a}(\Gb_{\psib}{}^\sigma{}_a
     -\Gb_{\psi}{}^\sigma{}_a)
     - \frac{2}{\tau}\Gb_\psi{}^{\sigma}{}_b(\Gb_A{}^{\sigma b}
     +\Gb_A{}^{\sigmab b})\nn\\
     &+2e^{-\varphi}\sg_f{}^e\Gb_{a A e}\Gb_\psi{}^{fa}
     -4\sg_{de}\Gb_A{}^d{}_b\Gb^{(\sigma e) b}\,,\nn\\
   \end{aligned}\\
  &\begin{aligned}
     (\Gamma\Gamma)_{\ul{\psi} A}:=&
     \frac{2}{\tau}\Gb^{\sigmab}{}_A{}^{a}(\Gb_{\psi}{}^{\sigmab}{}_a
     -\Gb_{\psib}{}^{\sigmab}{}_a)
     - \frac{2}{\tau}\Gb_{\psib}{}^{\sigmab}{}_b
     (\Gb_A{}^{\sigmab b}+\Gb_A{}^{\sigma b})\nn\\
     &+2e^{-\varphi}\sg_f{}^e\Gb_{a A e}\Gb_{\psib}{}^{fa}
     -4\sg_{de}\Gb_A{}^d{}_b\Gb^{(\sigmab e) b}\,,\nn\\
   \end{aligned}\\
  &(\cancel{\Gamma\Gamma}):=-\Gb^{bac}\Gb_e{}^d{}_c
  \left[\sg_{ad}(\delta_b^e-\sg_b{}^e)-\sg_{bd}\sg_a{}^e\right]\nn\,,\\
  &(\Gamma\Gamma)^{\theta\theta}:=
  -\mathbullet{\square}(\epsilon+\mathbullet{\epsilon})
  - \sg^{\theta\theta}\frac{1+\cos\theta^2}{\sin\theta^2}
  -\sg^a{}_b\sg^c{}_d\Gb_a{}^b{}_e\Gb_c{}^{de} \nn\\
  &\quad\quad2\cot\theta\sg^a{}_b\sg^{c\theta}\Gb_a{}^b{}_c
  + \frac{4}{\sg^{\theta\theta}}\Gb^{\theta\theta}{}_c
  \left(\sg^a{}_b\Gb_a{}^{bc}-\cot\theta g^{c\theta}\right)\nn\\	
  &\quad\quad +\frac{1}{\cosh h_\times}
  (\mathbullet{\square}\cosh h_\times
  + 2\nabla_a\cosh h_\times\nabla^ah_+)\nn\\
  &\quad\quad+\nabla_a h_+\nabla^a h_+ \nn\\
  &(\Gamma\Gamma)^{\theta\phi}:=-\sg_{\theta\phi}\coth h_\times
  \left[2\Gb_a{}^\theta{}_b\Gb^{a\phi b}
    +\sg^{\theta\phi}
    \mathbullet{\square}(\epsilon+\mathbullet{\epsilon})\right.\\
    &\left.- 4\sg^b{}_c\Gb^{(\theta\phi)}{}_a\Gb_b{}^{ca}
    + \sg^{\theta\phi}\sg^b{}_c\sg^d{}_e\Gb_b{}^c{}_f\Gb_d{}^{ef}
    - \sg^{\theta\phi}\nabla_a h_\times\nabla^a h_\times\right]\nn\,,
\end{align}
and~$\tilde{\mathcal{R}}_{ab}$ is an auxiliary tensor we introduce
simply to make expressions shorter,
\begin{align}\label{Rtilde}
 \tilde{\mathcal{R}}_{ab} := \mathcal{R}_{ab} + \nabla_{(a}F_{b)}+W_{ab}\,,
\end{align}
$\mathcal{R}_{ab}$ being the reduced Ricci tensor defined as,
\begin{align}
 \mathcal{R}_{ab} := R_{ab} - \nabla_{(a}Z_{b)}-W_{ab}\,,
\end{align}
where~$F^a$ are the gauge source functions and~$W_{ab}$ represents a a
generic constraint addition, by which we mean any expression
homogeneous in the constraints. We do not expect the reader to be
enlightened by these long expressions. However we choose to present
them nonetheless for the sake of completeness and in order to
highlight the fact that the information contained in them, together
with~\eqref{eqn:ConnectionComps} and the help of computer algebra,
provides a fairly quick way to write the EFEs in terms of derivatives
of the~$10$ metric functions. We will only be working in vacuum, so
all the components of the reduced Ricci tensor~$\mathcal{R}_{ab}$ are
zero. In order to get our final equations we only have to pick the
gauge source functions~$F^a$ and the constraint addition~$W_{ab}$. We
have seen in~\cite{DuaFenGas22} that the special interplay between
gauge choice and constraint addition can be exploited to turn the EFEs
into a system of~$8$ ugly equations with a natural number~$p$ and~$2$
good equations. These choices are highly non-unique as only the first
few orders of the solutions to these equations are concerned. Here we
want to find a choice that ensures this number of uglies and goods,
while preventing the appearance of logs up to order~$p$, and then show
the final form of the EFEs in the most convenient way.

\subsection{Constraints} 

Up to gauge fixing, the constraints~$Z^a:=\Gb^a + F^a$ are defined by
the following,
\begin{align}
  &\Gb^\sigma = \frac{2e^{-\varphi}}{\tau}\nabla_{\ul{\psi}}\mathcal{C}^R_+
  -\frac{2e^{-\varphi}}{\mathring{R}}\nabla_{\psi}\mathring{R}
  -\frac{\mathcal{C}_A}{\tau}\sD^A \mathcal{C}^R_+
  + \sD^A\mathcal{C}_A^+ \,,\nn\\
  &\Gb^{\ul{\sigma}} = -\frac{2e^{-\varphi}}{\mathring{R}}\nabla_{\ul{\psi}}
  \mathring{R}-\frac{2e^{-\varphi}}{\tau}\nabla_{\psi}\mathcal{C}^R_-
  +\frac{\mathcal{C}_A}{\tau}\sD^A \mathcal{C}^R_-
  +\sD^A\mathcal{C}_A^- \,,\nn\\
  &\Gb^A = \mathring{R}\left(\sg^{AB}\sD^Cq_{BC}
  -\frac{1}{2}\sg^{BC}\sD^Aq_{BC}\right)+\sD^A\varphi\nn\\
  &-\frac{\sg^{AB}}{\tau}\left(\nabla_{\ul{\psi}}
  \mathcal{\mathcal{C}}^+_B+\nabla_{\psi}\mathcal{C}^-_B
  -\mathcal{C}_B\frac{\nabla_{\ul{\psi}}\mathcal{C}^R_+
  -\nabla_{\psi}\mathcal{C}^R_-}{\tau}\right)\,,
\end{align}
where the only terms that contribute to leading order are the ones
proportional to bad derivatives of~$\cpr$, $\Rc$
and~$\mathcal{\mathcal{C}}^+_B$, respectively. This means that we can
write,
\begin{align}
  &Z^\sigma \simeq \nabla_{\ul{\psi}}\mathcal{C}^R_+ \,,\nn\\
  &Z^{\ul{\sigma}} \simeq
  -\frac{2}{\mathring{R}}\nabla_{\ul{\psi}}\mathring{R} \,,\nn\\
  &Z^A \simeq
  -\frac{\sg^{AB}}{2}\nabla_{\ul{\psi}}\mathcal{\mathcal{C}}^+_B\,.
\end{align}
The fact that, to leading order, each constraint is essentially a bad
derivative of a metric function will allow us to use constraint
addition for two separate purposes. The first is to turn~$4$ out
of~$10$ equations into uglies with a natural number~$p$, which is done
by adding constraints in such a way that they will contribute to
second order in~$R^{-1}$. Later in this work, we will see that bad
derivatives of~$\cpr$, $\Rc$ and~$\mathcal{\mathcal{C}}^+_A$ give rise
to formally singular terms that create problems in numerical
implementations. We will see that these terms appear at third order
in~$R^{-1}$ in some of the equations and that we can get rid of them
by constraint addition. For this reason, we will separate~$W_{ab}$
into a leading and a subleading contributions, $W_{ab}^{(1)}$
and~$W_{ab}^{(2)}$, respectively,
\begin{align}
  W_{ab}=W_{ab}^{(1)}+W_{ab}^{(2)}\,.
\end{align}
In this section, because we are only worried about establishing the
leading asymptotics of the metric fields and not the existence of
formally singular terms, we will only fix~$W_{ab}^{(1)}$ and
leave~$W_{ab}^{(2)}$ free for the moment.

\subsection{Gauge} 

We introduce the Cartesian harmonic gauge defined by~$F^a=\ring{F}^a$
with,
\begin{align}\label{eqn:GaugeSourcesForCartesianHarmonic}
  \ring{F}^a =g^{bc}
  \Gamma[\tilde{\nabla},\mathbullet{\nabla}]_b{}^a{}_c\,,
\end{align}
where~$\Gamma[\tilde{\nabla},\mathbullet{\nabla}]_{a}{}^{b}{}_{c}$ are
given functions of the coordinates. For conciseness we will use
explicitly standard spherical polar coordinates as shell coordinates
but in the code a different choice could be needed. For this
reason,~$\Gamma[\tilde{\nabla},\mathbullet{\nabla}]_b{}^a{}_c$ are
just the Christoffel symbols of Minkowski spacetime in polar
coordinates. Explicitly,
\begin{align}
  &\ring{F}^\sigma = \frac{2e^{-\epsilon}}{R}\cosh h_+\cosh h_\times
  + \mathcal{C}^+_A\ring{F}^A\,,\nn\\
  &\ring{F}^{\sigmab} = -\frac{2e^{-\epsilon}}{R}\cosh h_+\cosh h_\times
  + \mathcal{C}^-_A\ring{F}^A\,,\nn\\
  &\ring{F}^\theta = \frac{\cot \theta}{\Rc^2}e^{h_+}\cosh h_\times
  -\frac{2}{R}\sg^{R\theta}\,,\nn\\
  &\ring{F}^\phi =-\frac{2\cot\theta}{\sin \theta \Rc^2}
  \sinh h_\times-\frac{2}{R}\sg^{R\phi}\,.
\end{align}
We choose a gauge that is Cartesian harmonic to leading order,
with a higher order correction~$\check{F}^a$ which we specify later,
\begin{align}\label{eqn:gauge}
  &F^\sigma =\frac{2}{\Rc}\cosh h_+\cosh h_\times
  + \mathcal{C}^+_A\ring{F}^A + \check{F}^\sigma\nn\,,\\
  &F^{\sigmab} = -\frac{2}{\Rc}\cosh h_+\cosh h_\times
  + \mathcal{C}^-_A\ring{F}^A + \check{F}^{\sigmab}\,,\nn\\
  &F^\theta =\frac{\cot \theta}{\Rc^2}e^{h_+}\cosh h_\times
  -\frac{2}{\Rc}\sg^{R\theta}  + \check{F}^\theta\,,\nn\\
  &F^\phi =-\frac{2\cot \theta}{\sin \theta \Rc^2}\sinh h_\times
  -\frac{2}{\Rc}\sg^{R\phi} + \frac{\check{F}^\phi}{\sin^2\theta}\,.
\end{align}
Notice that we in its present form,~\eqref{eqn:gauge} does not have
any explicit~$\epsilon$ or~$R$. We can do this because it does not
require changing the fact that the gauge is Cartesian harmonic to
leading order and we choose to do it because the final expressions
turn out to be simpler if these objects are replaced by~$\Rc$.

\subsection{Ugly equations with~$p$}

In order to turn a wave equation into an ugly, we only need to
consider the leading order contributions of the various terms. In
total we have~$4$ constraints and~$4$ gauge source functions we are
free to add and choose, respectively. Each of these can be used to
turn one metric function into an ugly. In other words, this freedom
allows us to write~$8$ out of~$10$ equations as uglies. In the
following we explain how to do this for each of them.

\paragraph*{Radial coordinate light speed~$\cpr$:} The equation
for~$\cpr$ can be written as,
\begin{align}\label{eqn:CpR}
  &\mathbullet{\square} \mathcal{C}_+^R = e^{-\varphi}\psi^a\nabla_\psi F_a
  + e^{-\varphi}W_{\psi\psi} + (\Gamma\Gamma)_{\psi\psi}\,.
\end{align}
Plugging~\eqref{eqn:gauge} in the first term on the RHS we get that,
\begin{align}
  e^{-\varphi}\psi^a\nabla_\psi F_a = -\frac{2}{\Rc^2}\nabla_\psi\Rc
  -\frac{1}{\tau}F^\sigma\nabla_{\psib}\cpr + o^+(R^{-2})\,.
\end{align}
The notation~$f=o^+(h)$ means
\begin{align}
  \exists \epsilon>0 :
  \lim_{R\rightarrow\infty} \frac{f}{hR^{-\epsilon}}=0\,.
\end{align}
or in other words, fall off faster than~$f=o(h)$; more
precisely,~$o^+(h)=o(hR^{-\epsilon})$. Whenever it is unambiguous, we
will use~$\simeq$ to denote the presence of these higher order terms
because it makes expressions shorter. Using computer algebra, it is
possible to show that,
\begin{align}
  (\Gamma\Gamma)_{\psi\psi} =
  \frac{2e^{-\varphi}}{\tau^2}(\nabla_{\psib}\cpr)^2
  + \frac{2e^{-\varphi}}{\Rc^2}(\nabla_\psi\Rc)^2+o^+(R^{-2})\,.
\end{align}
As was said above, the constraint~$Z^\sigma$ is essentially a bad
derivative of~$\cpr$ to leading order and hence it can be used to make
sure that~\eqref{eqn:CpR} is an ugly by changing the RHS to
satisfy~\eqref{eqn:superugly2}. With this in mind we choose,
\begin{align}
  W_{\psi\psi}^{(1)}
  = Z^\sigma e^{\varphi}\left(\frac{p}{\Rc}\nabla_\psi \Rc
  - \frac{1}{\tau}\nabla_{\psib}\cpr\right)\,,
\end{align}
so that the wave equation for~$\cpr$ can be written as,
\begin{align}
  \mathbullet{\square}\cpr=
  \frac{2(p+1)e^{-\varphi}}{\tau\Rc}\nabla_\psi\Rc\nabla_{\psib}\cpr
  + \mathcal{N}_{\cpr}\,.
\end{align}
This means that~$\cpr$ is now an ugly with natural number~$p$.

\paragraph*{Radial coordinate light speed~$\cmr$:} The equation
for~$\cmr$ is the following,
\begin{align}\label{CmR}
  &\mathbullet{\square} \mathcal{C}_-^R
  = -e^{-\varphi}\psib^a\nabla_{\psib} F_a
  - e^{-\varphi}W_{\psib\psib} + (\Gamma\Gamma)_{\psib\psib}\,.
\end{align}
where the first term on the RHS, to leading order, behaves as,
\begin{align}
  -e^{-\varphi}\psib^a\nabla_{\psib} F_a \simeq
  -\nabla_{\psib}\check{F}^{\sigmab}
  +\frac{1}{\Rc}\nabla_{\psib}\cmr+ \frac{2}{\Rc^2}\,,
\end{align}
and the second as,
\begin{align}
  (\Gamma\Gamma)_{\psib\psib} \simeq -\frac{1}{2}(\nabla_{\psib}h_+)^2
  -\frac{1}{2}(\nabla_{\psib}h_\times)^2 - \frac{2}{\Rc^2}\,.
\end{align}
Because none of the constraints contains a bad derivative of $\cmr$,
we cannot use constraint addition to turn this variable into an
ugly. So we choose,
\begin{align}
  W_{\psib\psib}^{(1)} = 0\,.
\end{align} 
However, a bad derivative of~$F^{\sigmab}$ contributes to leading
order, so we can choose the gauge in order to get the asymptotic
behavior that we are looking for. In order to
get~\eqref{eqn:superugly2}, we need to make sure that our gauge choice
satisfies the following condition,
\begin{align}\label{FsigmabCondition}
  \nabla_{\psib}\check{F}^{\sigmab} \simeq \frac{1}{2}(\nabla_{\psib} h_+)^2
  + \frac{1}{2}(\nabla_{\psib} h_\times)^2
  - \frac{p}{\Rc}\nabla_{\psib}\mathcal{C}_-^R\,.
\end{align}
We do that by separating~$\check{F}^{\sigmab}$ in two parts,
\begin{align}
  \check{F}^{\sigmab} = \frac{1}{\Rc}\check{F}^{\sigmab}_1
  - \frac{p}{\Rc}(1+\mathcal{C}_-^R) \,,
\end{align}
where~$\check{F}^{\sigmab}_1$ is a function we will call \textit{gauge
  driver} since its purpose will be to drive the asymptotics of one of
the gauge source functions to a preassigned value at null
infinity. The condition~\eqref{FsigmabCondition} implies that,
\begin{align}\label{Fsigmab1Condition}
  \frac{1}{\Rc}\nabla_{\psib}\check{F}^{\sigmab}_1
  \simeq
  \frac{1}{2}(\nabla_{\psib} h_+)^2
  + \frac{1}{2}(\nabla_{\psib} h_\times)^2 \,,
\end{align}
which cannot be simply integrated. As other derivatives of the
function~$\check{F}^{\sigmab}_1$ will show up in other equations, it
is possible that these will interfere with the principal part of those
equations potentially spoiling hyperbolicity. The way to get around
this is to treat~$\check{F}^{\sigmab}_1$ as another evolved variable
and choosing a wave equation for it to satisfy that forces the
asymptotic condition~\eqref{Fsigmab1Condition}. As this equation will
not be an ugly, we dedicate to its treatment its own subsection at the
end of this section.

\paragraph*{Determinant of the metric in the~$T$-$R$ plane~$\varphi$:}
The equation for~$\varphi$ is,
\begin{align}
  \mathbullet{\square} \varphi = \frac{2e^{-\varphi}}{\tau}
  \psi^{(a}\psib^{b)}(\nabla_a F_b + W_{ab})+(\Gamma\Gamma)_{\psi\ul{\psi}}\,.
\end{align}
Since no constraint contains any bad derivative of~$\varphi$, we
choose,
\begin{align}
  W_{\psi\psib}^{(1)} =W_{\psib\psi}^{(1)} = 0\,,
\end{align}
and we must rely on gauge fixing to turn $\varphi$ into an ugly. To
leading order we have,
\begin{align}
  \frac{2e^{-\varphi}}{\tau}\psi^{(a}\psib^{b)}\nabla_a F_b
  \simeq \frac{1}{2}\nabla_{\psib}\check{F}^\sigma
  +\frac{1}{\Rc}\nabla_{\psib}\varphi+\frac{2}{\Rc^2}\,,
\end{align}
and,
\begin{align}
  (\Gamma\Gamma)_{\psi\ul{\psi}} \simeq -\frac{2}{\Rc}\,.
\end{align}
In order to obtain~\eqref{eqn:superugly2}, we must only guarantee
that~$F^\sigma$ satisfies,
\begin{align}
  \nabla_{\psib}\check{F}^\sigma \simeq
  \frac{2p}{\Rc}\nabla_{\psib}\varphi\,,
\end{align}
so we make the explicit choice,
\begin{align}
  \check{F}^\sigma = \frac{2p}{\Rc}(e^{\varphi}-1)\,.
\end{align}
Finally, we can write,
\begin{align}
  \mathbullet{\square}\varphi=\frac{2(p+1)e^{-\varphi}}{\tau\Rc}
  \nabla_\psi\Rc\nabla_{\psib}\varphi
  + \mathcal{N}_{\varphi}\,.
\end{align}

\paragraph*{Angular coordinate light speeds $\mathcal{C}^+_A$:} The
variables~$\mathcal{C}^+_A$ with wave equations,
\begin{align}
  \mathbullet{\square} \mathcal{C}^+_A = -2e^{-\varphi}\sg_A{}^{(a}\psi^{b)}
  (\nabla_aF_b+W_{ab})+(\Gamma\Gamma)_{\psi A}\,,
\end{align}
need to be rescaled in order to be classified as uglies, and the same
goes for~$\mathcal{C}^-_A$. That is done by defining the new
variables,
\begin{align}\label{Chat}
  \Rc\hat{\mathcal{C}}_A^\pm = \mathcal{C}_A^\pm\,,
\end{align}
and requiring that~$\hat{\mathcal{C}}_A^\pm$ behave as ugly fields
instead. This rescaling yields,
\begin{align}\label{ChatRescale}
  \mathbullet{\square}\mathcal{C}_A^\pm
  \simeq  \Rc\mathbullet{\square} \hat{\mathcal{C}}^\pm_A
  -\frac{2e^{-\varphi}}{\tau}\nabla_\psi
  \Rc\nabla_{\psib}\hat{\mathcal{C}}_A^\pm \,. 
\end{align}
Also, it can be seen that,
\begin{align}
  -2e^{-\varphi}\sg_A{}^{(a}\psi^{b)}\nabla_aF_b
  + (\Gamma\Gamma)_{\psi A}
  \simeq 2\nabla_{\psib}\hat{\mathcal{C}}_A^+\,,
\end{align}
so a choice of constraint addition we can use to
turn~$\hat{\mathcal{C}}_A^+$ into an ugly is,
\begin{align}
  W_{\psi A}^{(1)} = 2Z^A\left(p\nabla_\psi \mathring{R}
  - 2e^\varphi\right)\,.
\end{align}
Putting all this together we find,
\begin{align}
  \mathbullet{\square}\hat{\mathcal{C}}_A^+
  =  \frac{2(p+1)e^{-\varphi}}{\tau\Rc}\nabla_\psi
  \Rc\nabla_{\psib}\hat{\mathcal{C}}_A^+
  + \mathcal{N}_{\hat{\mathcal{C}}_A^+}\,. 
\end{align}

\paragraph*{Angular coordinate light speeds $\mathcal{C}^-_A$:} The
wave equation satisfied by $\mathcal{C}^-_A$ is,
\begin{align}
  \mathbullet{\square} \mathcal{C}^-_A
  = -2e^{-\varphi}\sg_A{}^{(a}\psib^{b)}
  (\nabla_aF_b+W_{ab})+(\Gamma\Gamma)_{\psib A}\,.
\end{align}
Rescaling~$\mathcal{C}^-_A$ as in~\eqref{Chat}
yields~\eqref{ChatRescale}. Since none of the constraints contain bad
derivatives of~$\hat{\mathcal{C}}_A^-$, we set,
\begin{align}
	\sg_A{}^{a}W_{(a\psib)}^{(1)}=0\,,
\end{align}
and we must now use our last~$2$ free gauge source functions~$F_A$. We
have that,
\begin{align}
  -2e^{-\varphi}\sg_A{}^{(a}\psib^{b)}&\nabla_aF_b
  \simeq 2\nabla_{\psib}\hat{\mathcal{C}}_A^-
  -\Rc^2\mbn_{\psib}\check{F}^A\\
  &+\cot\theta\delta_{\theta A}(2-\nabla_{\psib}h_+)
  +2\cos\theta\delta_{\phi A} \nabla_{\psib}h_\times\nn\,,
\end{align}
and,
\begin{align}
  (\Gamma\Gamma)_{\psib A} \simeq
  -\cot\theta\delta_{\theta A}(2-\nabla_{\psib}h_+)
  -2\cos\theta\delta_{\phi A} \nabla_{\psib}h_\times\,.
\end{align}
So the condition we need~$F^A$ to satisfy is,
\begin{align}
  \Rc^2\mbn_{\psib}\check{F}^A
  \simeq (1-p)\nabla_{\psib}\hat{\mathcal{C}}_A^-\,,
\end{align}
and one possible choice is,
\begin{align}
  \check{F}^A=
  \frac{1-p}{\Rc^2}\hat{\mathcal{C}}_A^-\,.
\end{align}
So we can write,
\begin{align}
  \mathbullet{\square}\hat{\mathcal{C}}_A^-
  =  \frac{2(p+1)e^{-\varphi}}{\tau\Rc}\nabla_\psi
  \Rc\nabla_{\psib}\hat{\mathcal{C}}_A^-
  + \mathcal{N}_{\hat{\mathcal{C}}_A^-}\,. 
\end{align}

\subsection{Ugly equation with~$p=1$}

\paragraph*{Determinant of the metric in the~$\theta$-$\phi$
  plane~$\epsilon$ or inverse areal radius~$\Rc^{-1}$:} There is only
one more metric function that we can possibly hope to turn into an
ugly field with natural number~$p$ with this method. The
field~$\epsilon$ satisfies the following equation,
\begin{align}
 \mathbullet{\square} (\epsilon + \mathbullet{\epsilon}) =
 -\sg^{ab}\nabla_aF_b - \cancel{W} + (\cancel{\Gamma\Gamma})\,,
\end{align}
and we want to deal with it a little differently. Instead of
evolving~$\epsilon$, we want to evolve~$\Rc^{-1}$. Notice that this
new variable is necessarily an ugly, since both good and bad
derivatives of it must improve, even if~$\epsilon$ is just a good
field. To see this we can expand the exponential in the definition
of~$\Rc$,
\begin{align}\label{expansionRc}
  \Rc^{-1} = \frac{1}{R}-\frac{\epsilon_{1,0}(\psi^*)}{2R^2}+o^+(R^{-2})\,,
\end{align}
where~$\epsilon_{1,0}(\psi^*)$ is a function that does not vary along
outgoing null curves. To leading order, the
quantities~$\mathbullet{\square}(\epsilon + \mathbullet{\epsilon})$
and~$\mathbullet{\square}\Rc^{-1}$ are related in the following way,
\begin{align}\label{boxepsToboxRc}
  \mathbullet{\square}(\epsilon + \mathbullet{\epsilon})\simeq
  2\Rc\mathbullet{\square}\Rc^{-1} - \frac{4e^{-\varphi}}{\tau}
  \nabla_\psi\Rc\nabla_{\psib}\Rc^{-1}\,.
\end{align}
Also, we have that,
\begin{align}
  -\sg^{ab}\nabla_aF_b \simeq
  - \frac{2\Rc}{\tau}F^\sigma\nabla_{\psib}\Rc^{-1}\,,
\end{align}
and,
\begin{align}
	(\cancel{\Gamma\Gamma}) =o^+(R^{-1}) \,.
\end{align}
Setting the constraint addition~$\cancel{W}$ to be,
\begin{align}
  \slashed{W}^{(1)} =
  Z^{\sigmab}\frac{2(\bar{p}-1)}{\tau\Rc}\nabla_\psi \Rc\,,
\end{align}
we get that~$\Rc^{-1}$ is an ugly with natural number~$\bar{p}$,
\begin{align}
  \mathbullet{\square}\Rc^{-1} =  \frac{2(\bar{p}+1)
    e^{-\varphi}}{\tau\Rc}\nabla_\psi \Rc\nabla_{\psib}\Rc^{-1}
  + \mathcal{N}_{\Rc^{-1}}\,,
\end{align}
where we introduced the natural number~$\bar{p}$ instead of~$p$
because we want~$\Rc^{-1}$ to behave differently from the other
fields.

\paragraph*{If~$\epsilon$ is a good, then the SNFs of~$\Rc^{-1}$
  are~$O(R^{-4})$:} This choice of constraint addition yields a wave
equation for~$\epsilon$ of the type,
\begin{align}\label{waveeqeps}
	\mathbullet{\square}\epsilon =
        \frac{2\bar{p}e^{-\varphi}}{\tau\Rc}\nabla_\psi
        \Rc\nabla_{\psib}\epsilon + \mathcal{N}_\epsilon\,.
\end{align}
Plugging~\eqref{waveeqeps} in~\eqref{boxepsToboxRc} then gives,
\begin{align}
  \mathbullet{\square}\Rc^{-1} =
  \frac{2(\bar{p}+1)e^{-\varphi}}{\tau\Rc}
  \nabla_\psi \Rc\nabla_{\psib}\Rc^{-1}
  + \frac{\bar{p}-1}{\Rc^3}+o^+(R^{-3})\,,
\end{align}
or equivalently,
\begin{align}
	\mathcal{N}_{\Rc^{-1}}\simeq\frac{\bar{p}-1}{\Rc^3}\,.
\end{align}
This means that if we choose~$\bar{p}=1$, we can ensure that the
stratified null forms in the wave equation for~$\Rc^{-1}$ are one
order better than what would normally be expected, $o^+(R^{-3})$
rather than~$o^+(R^{-2})$. According to~\eqref{waveeqeps}, this choice
makes~$\epsilon$ become a good field, while~$\Rc^{-1}$ remains an ugly
by construction, see~\eqref{expansionRc}. It is worth pausing here a
moment to analyze this choice from the point of view of the peeling
property. The strategy employed in~\cite{DuaFenGas22} to ensure that
the components of the Weyl tensor peel was to make~$8$ metric
functions be ugly with a natural number~$p$ that guarantees no logs
are generated up to that order. Here, by choosing~$\epsilon$ to be a
good, we are forced to have~$\Rc^{-1}$ be an ugly with~$\bar{p}=1$, so
by what we know about the subleading asymptotics of generic ugly
fields, we could expect logarithmically divergent terms to appear from
second order near null infinity. However~$\Rc^{-1}$ is not a generic
ugly. It is built purely from~$\epsilon$. Now we also know that good
fields cannot generate logs, they can merely inherit them. So,
despite~$\Rc^{-1}$ being an ugly with a fixed~$\bar{p}=1$, it cannot
generate logs at any order down in the expansion. In other words,
for~$p$ sufficiently large ($p=7$, see~\cite{DuaFenGas22}), the Weyl
tensor necessarily still peels (with the strict
  assumptions~\eqref{assumptionTimeder} on initial data).

\subsection{Good equations} 

\paragraph*{$h_+$ and~$h_\times$:} The wave equations satisfied by~$h_+$
and~$h_\times$ are,
\begin{align}
  &\mathbullet{\square}h_+ =
  -\frac{2}{\sg^{\theta\theta}}\left[(\nabla
    F)^{\theta\theta}+W^{\theta\theta}\right] +
  (\Gamma\Gamma)^{\theta\theta}\nn\,,\\
  &\mathbullet{\square}h_\times
  = -2\sg_{\theta\phi}\coth h_\times\left[(\nabla
    F)^{\theta\phi}+W^{\theta\phi}\right] +
  (\Gamma\Gamma)^{\theta\phi}\,.
\end{align}
We can see that,
\begin{align}
  &-\frac{2}{\sg^{\theta\theta}}(\nabla F)^{\theta\theta}
  + (\Gamma\Gamma)^{\theta\theta}= \frac{1}{\Rc}\nabla_{\psib}h_+
  + o^+(R^{-2})\,,\\
  &-2\sg_{\theta\phi}\coth h_\times(\nabla F)^{\theta\phi}
  + (\Gamma\Gamma)^{\theta\phi}=  \frac{1}{\Rc}\nabla_{\psib}h_\times
  +o^+(R^{-2})\,,\nn
\end{align}
so, according to~\eqref{gEqShort} we do not need to add constraints to
get the behavior we are looking for. This means that,
\begin{align}
  W^{(1)\theta\theta}=W^{(1)\theta\phi}=0\,,
\end{align}
and we can write both equations as standard good equations,
\begin{align}
  &\mathbullet{\square}h_+ =
  \frac{2e^{-\varphi}}{\tau\Rc}\nabla_\psi\Rc\nabla_{\psib}h_+
  + \mathcal{N}_{h_+}\nn\,,\\
  &\mathbullet{\square}h_\times =
  \frac{2e^{-\varphi}}{\tau\Rc}\nabla_\psi\Rc\nabla_{\psib}h_\times
  + \mathcal{N}_{h_\times}\,.
\end{align}

\subsection{Gauge driver}

\paragraph*{Gauge driver~$\check{F}^{\sigmab}_1$:} To make sure
that hyperbolicity remains untouched while the gauge source functions
satisfy~\eqref{FsigmabCondition} we need to
treat~$\check{F}^{\sigmab}_1$ as an eleventh evolved variable and
require it so satisfy a wave equation that forces the asymptotic
behavior~\eqref{FsigmabCondition}. Let us then consider the equation,
\begin{align}\label{waveeqF}
  \mathbullet{\square}\check{F}^{\sigmab}_1 =
  &\frac{2e^{-\varphi}}{\tau\Rc}\left[(p+1)\nabla_\psi
    \Rc\nabla_{\psib}\check{F}^{\sigmab}_1
    -\frac{p}{\Rc}\mathcal{H}\right]\,,
\end{align}
where~$2\mathcal{H}/\Rc^2:=(\nabla_{\psib} h_+)^2 + (\nabla_{\psib}
h_\times)^2$. We want to show that the wave equation~\eqref{waveeqF}
makes~$\check{F}^{\sigmab}_1$ satisfy~\eqref{FsigmabCondition}, so for
now we are only interested in the leading order terms. Therefore, we
can write
\begin{align}
  \nabla_\psi\nabla_{\psib}f_1\simeq
  -\frac{p}{R}\left[\nabla_{\psib}f_1-\mathcal{H}\right]\,,
\end{align}
where~$f:=\Rc\check{F}^{\sigmab}_1$. Adding the stratified null
form~$\nabla_\psi \mathcal{H}$ to either side, which we can do because
SNFs do not influence the leading order behavior by definition, we get
an ODE for the leading behavior
of~$\nabla_{\psib}\check{F}^{\sigmab}_1-\mathcal{H}$,
\begin{align}
  \nabla_\psi\left[\nabla_{\psib}f_1-\mathcal{H}\right]\simeq
  -\frac{p+1}{R}\left[\nabla_{\psib}f_1-\mathcal{H}\right]\,.
\end{align}
This can be integrated to give the asymptotics we are looking for,
\begin{align}\label{f-hdecay}
  \nabla_{\psib}f_1-\mathcal{H} \simeq \frac{\alpha(\psi^*)}{R^{p}}\,,
\end{align}
as long as~$p$ is a natural number. The wave equation satisfied by the
gauge driver in~\cite{DuaFenGas22} is not exactly the same as this
one, but they coincide asymptotically. Therefore, the proof that the
variable~$\check{F}^{\sigmab}_1$ cannot introduce any logs in the
system up to order~$p$ goes through in exactly the same way and we
repeat it here. Note that we we have assumed that the gauge driver
satisfies the same requirements on initial data than the rest of the
metric fields namely~\eqref{assumptionTimeder}.

\section{The Einstein field equations in second order form}
\label{section:EFEsSecondOrder}

Based on the discussion from the last section, we summarize the
choices we have made for the constraint additions and gauge source
functions. Then we write the resulting second order equations
explicitly as concisely as possible. The choice of constraint addition
encoded by the tensor~$W^{(1)}_{ab}$ is the following,
\begin{align}\label{eqn:ConstAdd2}
  &W^{(1)}_{\psi\psi} =
  Z^\sigma e^{\varphi}\left(\frac{p}{\Rc}\nabla_\psi \Rc
  - \frac{1}{\tau}\nabla_{\psib}\cpr\right)\,,\nn\\
  &W^{(1)}_{\psi A} =
  2Z^A\left(p\nabla_\psi \mathring{R} - 2e^\varphi\right)\,,
\end{align}
with all other components set to zero. Notice that up to this point we
have not yet chosen~$W^{(2)}_{ab}$ for any of the equations and we do
not have to, since by construction this choice bears no influence on
the leading asymptotics of the metric functions. The choice of gauge
is,
\begin{align}\label{gaugefinal}
  &F^\sigma = \frac{2}{\Rc}\cosh h_+\cosh h_\times
  + \mathcal{C}^+_A\ring{F}^A + \frac{2p}{\Rc}(e^{\varphi}-1)\,,\nn\\
  &F^{\sigmab} = -\frac{2}{\Rc}\cosh h_+\cosh h_\times
  + \mathcal{C}^-_A\ring{F}^A - \frac{p}{\Rc}(1+\mathcal{C}_-^R)
  + \frac{1}{\Rc}\check{F}^{\sigmab}_1 \,,\nn\\
  &F^\theta = \frac{\cot \theta}{\Rc^2}e^{h_+}\cosh h_\times
  -\frac{2}{\Rc}\sg^{R\theta}
  + \frac{1-p}{\Rc^2}\hat{\mathcal{C}}_\theta^-\,,\nn\\
  &F^\phi = -\frac{2\cot \theta}{\sin \theta \Rc^2}\sinh h_\times
  -\frac{2}{\Rc}\sg^{R\phi}
  +\frac{1-p}{\sin^2\theta\Rc^2}\hat{\mathcal{C}}_\phi^-\,.
\end{align}

\paragraph*{Einstein field equations:} Here it pays off to change the
variables that usually appear inside exponentials in such a way that
we can clearly separate the leading~$\pm 1$ from the functions we wish
to evolve. This will allow us to rescale the variables directly
by~$\Rc$, as we will have to do later on. Let~$\tilde{\varphi}$ denote
field~$e^\varphi-1$, $\tilde{h}_+$ denote~$e^{h_+}-1$
and~$\tilde{h}_\times$ denote~$e^{h_\times}-1$. Then the wave
equations for these variables change like,
\begin{align}
  &\mathbullet{\square}_q\tilde{\varphi} =
  \tilde{\mathcal{N}}_\varphi(\tilde{\varphi}+1)
  -\frac{1}{\tilde{\varphi}+1}
  \nabla_a\tilde{\varphi}\nabla^a\tilde{\varphi}\,,\nn\\
  &\mathbullet{\square}_q\tilde{h}_+ =
  \tilde{\mathcal{N}}_{h_+}(\tilde{h}_++1)
  -\frac{1}{\tilde{h}_++1}
  \nabla_a\tilde{h}_+\nabla^a\tilde{h}_+\,,\nn\\
  &\mathbullet{\square}_q\tilde{h}_\times
  =\tilde{\mathcal{N}}_{h_\times}(\tilde{h}_\times+1)
  -\frac{1}{\tilde{h}_\times+1}
  \nabla_a\tilde{h}_\times\nabla^a\tilde{h}_\times\,.
\end{align}
Although this choice of variables changes the second order equations,
it does so in such a way that the leading order remains
unaffected. Hence good equations remain good, ugly ones remain ugly
and so on. Expanding the wave operator and using~\eqref{boxp} we can
write all~$11$ field equations as,
\begin{align}\label{finaleqs}
  &\mathbullet{\square}_p\cpr =
  \sigma^a\nabla_{\psi} F_a
  +e^{-\varphi}W_{\psi\psi}
  - \frac{2(p+1)e^{-\varphi}}{\tau\Rc}\nabla_\psi\Rc\nabla_{\psib}\cpr\nn\\
  &\quad\quad\quad\quad+(\Gamma\Gamma)_{\psi\psi} \,,\nn\\
  &\mathbullet{\square}_p\cmr =
  -\sigmab^a\nabla_{\psib} F_a +(\Gamma\Gamma)_{\psib\psib}
  -\frac{2(p+1)e^{-\varphi}}{\tau\Rc}\nabla_\psi\Rc\nabla_{\psib}\cmr\,,\nn\\
  &\mathbullet{\square}_p\tilde{\varphi} =
  \frac{2e^{-\varphi}}{\tau}\sigma^{(a}\psib^{b)}\nabla_a F_b
  + e^{-\varphi}(\Gamma\Gamma)_{\psi\psib}\nn\\
  &\quad\quad\quad
  -\frac{2(p+1)e^{-3\varphi}}{\tau\Rc}\nabla_\psi\Rc\nabla_{\psib}\tilde{\varphi}
  -e^{-\varphi}\nabla_a\tilde{\varphi}\nabla^a\tilde{\varphi}\nn\,,\\
  &\mathbullet{\square}_p\mathcal{\hat{C}}^+_A
  = -2e^{-\varphi}\sg_A^a\psi^b\nabla_{(a}F_{b)}
  + \frac{2e^{-\varphi}}{\tau}\nabla_{\psib}\Rc\nabla_\psi\mathcal{\hat{C}}^+_A
  + (\Gamma\Gamma)_{\psi A}\nn\\
  &\quad\quad\quad\quad-2e^{-\varphi}\sg_A^aW_{\psi a}
  -\frac{2pe^{-\varphi}}{\tau\Rc^2}
  \nabla_\psi\Rc\nabla_{\psib}\mathcal{\hat{C}}^+_A\nn\\
  &\quad\quad\quad\quad
  -\mathcal{\hat{C}}^+_A\mathbullet{\square}\Rc
  -2\slashed{\nabla}_a\mathcal{\hat{C}}^+_A\slashed{\nabla}^a\Rc\,,\nn\\
  &\mathbullet{\square}_p\mathcal{\hat{C}}^-_A
  = -2e^{-\varphi}\sg_A^a\psib^b\nabla_{(a}F_{b)}
  + \frac{2e^{-\varphi}}{\tau}\nabla_{\psib}\Rc\nabla_\psi\mathcal{\hat{C}}^-_A
  + (\Gamma\Gamma)_{\psib A}\nn\\
  &\quad\quad\quad\quad
  -\frac{2pe^{-\varphi}}{\tau\Rc^2}
  \nabla_\psi\Rc\nabla_{\psib}\mathcal{\hat{C}}^-_A
  -\mathcal{\hat{C}}^-_A\mathbullet{\square}\Rc
  -2\slashed{\nabla}_a\mathcal{\hat{C}}^-_A\slashed{\nabla}^a\Rc\,,\nn\\
  &\mathbullet{\square}_1\Rc^{-1} =\frac{1}{2\Rc}\left[-\sg^{ab}\nabla_aF_b
    - \cancel{W} + (\cancel{\Gamma\Gamma})
    - \sg^{\theta\theta}(1+\cot\theta^2)\right]\nn\\
  &\quad\quad\quad\quad+ \frac{1}{\Rc}\nabla_a\Rc\nabla^a\Rc\,,\nn\\
  &\mathbullet{\square}_0 h_+ =
  -\frac{2e^{-h_+}}{\sg^{\theta\theta}}(\nabla F)^{\theta\theta}
  + e^{-h_+}(\Gamma\Gamma)^{\theta\theta} \nn\\
  &\quad\quad\quad\quad
  - \frac{2e^{-\varphi-2h_+}}{\tau\Rc}
  \nabla_\psi\Rc\nabla_{\psib}\tilde{h}_+-e^{-h_+}
  \nabla_a\tilde{h}_+\nabla^a\tilde{h}_+\,,\nn\\
  &\mathbullet{\square}_0 h_\times
  =  -\frac{2e^{-h_\times}}{\sg^{\theta\theta}}(\nabla F)^{\theta\phi}
  + e^{-h_\times}(\Gamma\Gamma)^{\theta\phi} \nn\\
  &\quad\quad\quad\quad- \frac{2e^{-\varphi-2h_\times}}{\tau\Rc}
  \nabla_\psi\Rc\nabla_{\psib}\tilde{h}_\times-e^{-h_\times}
  \nabla_a\tilde{h}_\times\nabla^a\tilde{h}_\times\,,\nn\\
  &\mathbullet{\square}_p\check{F}^{\sigmab}_1
  + \frac{2pe^{-\varphi}}{\tau\Rc^2}\mathcal{H}=0\,.
\end{align}
Only the LHSs of these equations can possibly contribute to leading
order and all of the RHSs are stratified null forms. This means that
the leading order asymptotics of each metric function, and therefore
whether it qualifies as a good or an ugly or otherwise, is determined
solely by the operator on the LHS. Therefore the~$11$ equations can be
written in the very concise form,
\begin{align}\label{finalwaveseqs}
  \mathbullet{\square}_{q} \phi = \mathcal{N}_\phi
  \quad,\quad
  \mathbullet{\square}_p\check{F}^{\sigmab}_1
  = -\frac{p+1}{\Rc}\mathcal{H}\,,
\end{align}
where~$\phi\in\{\mathcal{C}^R_{\pm},\hat{\mathcal{C}}^\pm_A,\varphi,
\Rc^{-1},h_+,h_\times\}$, $q=p$
for~$\phi\in\{\mathcal{C}^R_{\pm},\hat{\mathcal{C}}^\pm_A,\varphi\}$,
$q=1$ for~$\phi=\Rc^{-1}$ and~$q=0$ for~$\phi\in\{h_+,h_\times\}$. The
stratified null forms~$\mathcal{N}_\phi$ are the RHSs of the
corresponding equations in~\eqref{finaleqs}.

\paragraph*{Alternative form:} A less concise way to write the wave
equations that is more useful for what we aim to do in the next
section makes use of~\eqref{uEqLong},
\begin{align}\label{alternativeform}
  &\frac{1}{\Rc^{q+1}}\nabla_\psi
  \left[\Rc^{q+1}\nabla_{\psib}\phi\right]
  - \frac{\tau e^\varphi}{2}\slashed{\Delta}\phi
  =\tilde{\mathcal{N}}_\phi\,,\\
  &\frac{1}{\Rc^{p+1}}\nabla_\psi
  \left[\Rc^{p+1}\nabla_{\psib}\check{F}^{\sigmab}_1\right]
  - \frac{\tau e^\varphi}{2}\slashed{\Delta}\check{F}^{\sigmab}_1
  -\frac{p}{\Rc^2}\mathcal{H}
  =\tilde{\mathcal{N}}_{\check{F}^{\sigmab}_1}\,.\nn
\end{align}
This amounts to simply moving all the stratified null forms
in~$\mathbullet{\square}_q\phi$ to the RHS and
redefining~$\tilde{\mathcal{N}}_\phi$ so that all the principal terms
and those that contribute to leading order are on the LHS.

\section{First order reduction}
\label{section:Reduction}

In order to implement the EFEs in this framework numerically it is
useful to reduce them to a system of first order differential
equations. This is done by defining reduction variables, then
rescaling the evolved fields and compactifying the radial
coordinate. Although the equations in question are naturally more
numerous and complicated, most of this goes along the lines of the
work done in~\cite{GasGauHil19}.

\subsection{Picking the variables}

We begin by choosing the variables we will use in order to build our
first order system. As was said above, it pays off to avoid special
functions of variables, e.g. exponentials, in the equations because
they have a leading~$1$ that is not explicit. We leave a more detailed
explanation of this problem for the last section of this work. Let
$\phi$ denominate any of the following fields,
\begin{align}
 &\mathcal{C}_\pm^R\mp 1\,, \quad \mathcal{\hat{C}}^\pm_A\,, \quad
 e^{\varphi}-1\,, \quad \Rc^{-1}\,,  \nn\\
 & e^{h_+}-1\,,\quad e^{h_\times}-1, \quad \check{F}^{\sigmab}_1\,.
\end{align}
Moreover, let~$\phi_{,\psi}$, $\phi_{,\psib}$ and~$\phi_{,A}$
denote~$\nabla_\psi\phi$, $\nabla_{\psib}\phi$ and~$\sD_A\phi$,
respectively.

\subsection{Picking the equations}

The first order reduction is done by treating~$\phi_{,\psi}$,
$\phi_{,\psib}$ and~$\phi_{,A}$ as independent variables and the
second order differential equations as first order ones for these
variables. In order to relate them to the original variables~$\phi$,
we will choose one (or a combination) of the definitions as an
additional evolution equation and treat the others,
\begin{align}
  \phi_{,\psi} = \nabla_\psi\phi\,,\quad\phi_{,\psib}
  = \nabla_{\psib}\phi\,,\quad\phi_{,A} = \sD_A\phi\,,
\end{align}
as reduction constraints. After the reduction, where we had~$11$
independent variables, we end up with~$4$ additional variables per
original one so we will need~$55$ independent equations. The first set
of equations is naturally the wave equations~\eqref{finalwaveseqs},
which we can now write in terms of the reduction variables. This way
we have,
\begin{align}\label{waveeqNorescale}
  \frac{1}{\Rc^{q+1}}\nabla_\psi\left[\Rc^{q+1}\phi_{\psib}\right]
  - \frac{\tau e^\varphi}{2}\sD^A\phi_A =\tilde{\mathcal{N}}_\phi\,,
\end{align}
for the uglies and goods. Finally, the equation for the derivative
along~$\psib$ of the gauge driver reads,
\begin{align}\label{waveeqF1}
  \frac{1}{\Rc^{p+1}}\nabla_\psi
  \left[\Rc^{p+1}\check{F}^{\sigmab}_{1,\psib}\right]
  - \frac{\tau e^\varphi}{2}\sD^A\check{F}^{\sigmab}_{1,A}
  -\frac{p}{\Rc^2}\mathcal{H} =\tilde{\mathcal{N}}_{\check{F}^{\sigmab}_1}\,.
\end{align}
Replacing~$\phi$ with the appropriate fields, this gives~$11$ first
order differential equations for the variables~$\phi_{,\psib}$. The
second and third sets of equations will be given by the torsion-free
condition~$\mbn_a\nabla_b\phi=\mbn_b\nabla_a\phi$ contracted with the
vectors~$\psi^a$ and~$\psib^a$, and~$T^a$ and~$\sg_A{}^a$,
respectively. Let us take the first combination of vectors,
\begin{align}
  &\psi^a\psib^b\mbn_a\nabla_b\phi=\psi^a\psib^b\mbn_b\nabla_a\phi\,,\\
  \Rightarrow & \nabla_\psi\nabla_{\psib}\phi
  - (\mbn_\psi\psib)^a\nabla_a\phi =\nabla_{\psib}\nabla_{\psi}\phi
  - (\mbn_{\psib}\psi)^a\nabla_a\phi \,,\nn\\
  \Rightarrow &\nabla_\psi\phi_{,\psib} -\nabla_{\psib}\phi_{,\psi}
  = \frac{1}{\tau}\left(\mathcal{C}_{-,\psi}^R
  -\mathcal{C}_{+,\psib}^R\right)
  \left(\phi_{,\psi}-\phi_{,\psib}\right)\,.\nn
\end{align}
We already have an evolution equation for~$\phi_{,\psib}$
in~\eqref{waveeqNorescale}, so we want~\eqref{notorsionNorescale1} to
be one for~$\phi_{,\psi}$. To do this, we expand the LHSs
of~\eqref{waveeqNorescale} and~\eqref{waveeqF1} and plug them
in~\eqref{notorsionNorescale1} to get,
\begin{align}\label{notorsionNorescale3}
  \nabla_{\psib}\phi_{,\psi} = &\frac{\tau e^\varphi}{2}\sD^A\phi_{,A}
  +\tilde{\mathcal{N}}_\phi -(q+1)\Rc\phi_{,\psib} (\Rc^{-1})_{,\psi} \nn\\
  &- \frac{1}{\tau}\left(\mathcal{C}_{+,\psib}^R
  -\mathcal{C}_{-,\psi}^R\right)\left(\phi_{,\psib}-\phi_{,\psi}\right)\,.
\end{align}
We are free to add any amount of the constraints we wish to these
equations and it is worth doing so in~\eqref{notorsionNorescale3} in
order to get rid of the bad derivative of~$\cpr$. We do this for
reasons we will explain later on when we discuss the existence of
formally singular terms. We add,
\begin{align}
\frac{e^\varphi}{2}Z^\sigma\left(\phi_{,\psib}-\phi_{,\psi}\right)\,,
\end{align}
so that~\eqref{notorsionNorescale3} becomes,
\begin{align}\label{notorsionNorescale1}
  \nabla_{\psib}&\phi_{,\psi} = \frac{\tau e^\varphi}{2}\sD^A\phi_{,A}
  +\tilde{\mathcal{N}}_\phi +(q+1)\Rc\phi_{,\psib} (\Rc^{-1})_{,\psi} \nn\\
  &- \frac{1}{\tau}\left(\mathcal{C}_{+,\psib}^R-\mathcal{C}_{-,\psi}^R
  -\frac{\tau e^\varphi}{2}Z^\sigma\right)
  \left(\phi_{,\psib}-\phi_{,\psi}\right)\,.
\end{align}
for all metric functions except the gauge driver and,
\begin{align}
  \nabla_{\psib}&\check{F}^{\sigmab}_{1,\psi}
  = \frac{\tau e^\varphi}{2}\sD^A\check{F}^{\sigmab}_{1,A}
  +\frac{p}{\Rc^2}\mathcal{H}
  +(p+1)\Rc\check{F}^{\sigmab}_{1,\psib} (\Rc^{-1})_{,\psi}\nn \\
  &- \frac{1}{\tau}\left(\mathcal{C}_{+\psib}^R
  -\mathcal{C}_{-\psi}^R
  -\frac{\tau e^\varphi}{2}Z^\sigma\right)
  \left(\check{F}^{\sigmab}_{1,\psib}
  -\check{F}^{\sigmab}_{1,\psi}\right)
  +\tilde{\mathcal{N}}_\phi\,.\nn
\end{align}
This gives us~$11$ additional equations. Following the same procedure
with the second combination of vectors we get and evolution equation
for~$\phi_{,A}$,
\begin{align}
  &\tau\nabla_{T}\phi_{,A}-\cpr\sD_A\phi_{,\psib}
  +\cmr\sD_A\phi_{,\psi} =\nn\\
  & \phi_{,\psi}\left(\frac{\cmr}{\tau}\tau_{,A}
  +\frac{\mathcal{C}_A}{\tau}\nabla_T\cmr
  +\nabla_{T}\mathcal{C}^+_A+\sD_A\cpr\right)\\
  &+\phi_{,\psib}\left(-\frac{\cpr}{\tau}\tau_{,A}
  -\frac{\mathcal{C}_A}{\tau}\nabla_T\cpr
  +\nabla_{T}\mathcal{C}^-_A-\sD_A\cmr\right)\,,\nn
\end{align}
where~$\nabla_T$ was used merely as shorthand
for~$\frac{1}{\tau}(\cpr\nabla_{\psib}-\cmr\nabla_\psi)$. Again we are
free to add any amount of the constraints to this equation and we
choose to add,
\begin{align}
  W^{(3)}_\phi=\frac{\cpr\Rc e^\varphi}{2\p_R \Rc}
  \p_R\sD_A(\Rc\phi)Z^{\sigmab}\,,
\end{align}
to the LHS, to get,
\begin{align}\label{notorsionNorescale2}
  &\tau\nabla_{T}\phi_{,A}+W^{(3)}_\phi-\cpr\sD_A\phi_{,\psib}
  +\cmr\sD_A\phi_{,\psi} =\nn\\
  & \phi_{,\psi}\left(\frac{\cmr}{\tau}\tau_{,A}
  +\frac{\mathcal{C}_A}{\tau}\nabla_T\cmr
  +\nabla_{T}\mathcal{C}^+_A+\sD_A\cpr\right)\\
  &+\phi_{,\psib}\left(-\frac{\cpr}{\tau}\tau_{,A}
  -\frac{\mathcal{C}_A}{\tau}\nabla_T\cpr
  +\nabla_{T}\mathcal{C}^-_A-\sD_A\cmr\right)\,.\nn
\end{align}
This is not done in order to eliminate formally singular terms, but to
make the proof of symmetric hyperbolicity of the final system of
equations more straightforward. This gives us another~$22$ equations
and it is valid for all the metric functions. We now have one equation
for each of the reduction variables, leaving us only with the task of
choosing equations that will relate these to the original
variables. We pick the equations,
\begin{align}\label{redConstraintNorescale}
  \phi_{,\psib} = \nabla_{\psib}\phi\,,
\end{align}
to serve that purpose and treat,
\begin{align}\label{RedConstraints}
  \phi_{,\psi} = \nabla_\psi\phi\,,\quad\phi_{,A}
  = \sD_A\phi\,,
\end{align}
as constraints that one needs to make sure are satisfied everywhere
when evolving the system numerically.

\subsection{Rescaling the variables}

We know from previous studies on the good-bad-ugly model that the
fields themselves have decay and their derivatives may enhance that
decay or keep it the same depending on if the derivative is good or
bad. Ideally, when implementing this model numerically, we would have
regular equations for variables that we expect to have regular
behavior and a finite value at null infinity. In this section we will
rescale the fields and the reduction variables with powers of~$\Rc$ to
find the best possible variables to evolve. Naturally we cannot
rescale the variable~$\Rc^{-1}$ by powers of~$\Rc$, so we will have to
treat this particular variable differently. We begin by rescaling all
the other variables in the following way,
\begin{align}\label{rescaling}
  &\Phi:=\Rc\phi\,,\quad\Phi_\psi:=\Rc\nabla_\psi\Phi\,,\nn\\
  &\Phi_{\psib}:=\nabla_{\psib}\Phi\,,\quad\Phi_A:=\sD_A\Phi\,.
\end{align}
Having in mind the asymptotics of goods and uglies derived
in~\cite{DuaFenGasHil21}, and of the gauge driver
in~\cite{DuaFenGas22}, we can expect this rescaling to give reduction
variables that asymptote to finite values or zero at null infinity in
all cases. For the variable~$\Rc^{-1}$ we introduce a different
rescaling,
\begin{align}\label{rescalingRc}
  &\rho_\psi:=-\Rc\left[\Rc^2(\Rc^{-1})_{,\psi}+1\right]\,,
  \quad\rho_{\psib}:=-\Rc^2(\Rc^{-1})_{,\psib}+1\,,\nn\\
  &\rho_A:=-\Rc^2(\Rc^{-1})_{,A}\,,
\end{align}
which can be written alternatively as,
\begin{align}\label{rescalingRc2}
  &\nabla_\psi\Rc=1+\frac{\rho_\psi}{\Rc}\,,
  \quad\nabla_{\psib}\Rc=-1+\rho_{\psib}\,,\nn\\
  &\sD_A\Rc=\rho_A\,.
\end{align} 
Plugging this into the equations derived in the last section,
\eqref{waveeqNorescale}, \eqref{notorsionNorescale1},
\eqref{notorsionNorescale2} and~\eqref{redConstraintNorescale}, we
find,
\begin{align}\label{EqsRescaled}
  &\frac{1}{\Rc^{q+1}}\nabla_\psi\left[\Rc^q\Phi_{\psib}\right]
  -\frac{\tau e^\varphi}{2\Rc}\sD^A\Phi_A = \mathcal{N}_\phi^{(1)}\,,\nn\\
  &\frac{1}{\Rc^2}\nabla_{\psib}\Phi_{\psi}
  -\frac{\tau e^\varphi}{2\Rc}\sD^A\Phi_A
  + \frac{q}{\Rc^2}\Phi_{\psib}
  = \mathcal{N}_\phi^{(2)}\,,\nn\\
  &\tau\nabla_T\Phi_A+W^{(3)}_\phi -\cpr\sD_A\Phi_{\psib}
  + \frac{\cmr}{\Rc}\sD_A\Phi_{\psi}
  = \mathcal{N}_\phi^{(3)}\Rc\,,\nn\\
  &\nabla_{\psib}\Phi -\Phi_{\psib}=0\,,
\end{align}
where we have packed all the stratified null forms in the definitions
of~$\mathcal{N}_\phi^{(1)}$, $ \mathcal{N}_\phi^{(2)}$
and~$\mathcal{N}_\phi^{(3)}$ on the RHS,
\begin{align}\label{SNFs1}
  & \mathcal{N}_\phi^{(1)} :=
  \tilde{\mathcal{N}}_\phi- \Phi\tilde{\mathcal{N}}_{\Rc^{-1}}
  +\frac{\Phi_\psi}{\Rc^3} (\rho_{\psib}-1)
  - \frac{\tau e^\varphi\Phi^A}{\Rc^2} \rho_A\,,\nn\\
  & \mathcal{N}_\phi^{(2)}:=
  \frac{1}{\tau\Rc}\left(\mathcal{C}_{+,\psib}^R
  -\mathcal{C}_{-,\psi}^R
  -\frac{\tau e^\varphi}{2}Z^\sigma\right)
  \left(\frac{\Phi_{\psi}}{\Rc}-\Phi_{\psib}\right) \nn\\
  -&\frac{q}{\Rc^3}\Phi_{\psib}\rho_\psi
  + \frac{2\Phi_\psi}{\Rc^3}(\rho_{\psib}-1)
  - \frac{\tau e^\varphi\Phi^A}{\Rc^2} \rho_A
  +\tilde{\mathcal{N}}_\phi -\Phi\tilde{\mathcal{N}}_{\Rc^{-1}}\,,\nn\\
  & \mathcal{N}_\phi^{(3)} =
  \frac{1}{\Rc}\Phi_{\psib}\left(-\frac{\cpr}{\tau}\tau_{,A}
  -\frac{\mathcal{C}_A}{\tau}\nabla_T\cpr+\nabla_{T}\mathcal{C}^-_A
  -\sD_A\cmr\right)\nn\\
  +&\frac{1}{\Rc^2}\Phi_{\psi}\left(2\cmr\rho_A
  +\frac{\cmr}{\tau}\tau_{,A}+\frac{\mathcal{C}_A}{\tau}\nabla_T\cmr
  +\nabla_{T}\mathcal{C}^+_A+\sD_A\cpr\right).
\end{align}
To obtain the equations for the gauge driver we only need to
modify~\eqref{EqsRescaled} slightly in order to include the RHS of the
third equation in~\eqref{finalwaveseqs},
\begin{align}\label{EqGaugeDriverRescaled}
  &\frac{1}{\Rc^{p+1}}\nabla_\psi\left[\Rc^p\Phi_{\psib}\right]
  -\frac{\tau e^\varphi}{2\Rc}\sD^A\Phi_A
  -\frac{p}{\Rc^2}\mathcal{H}
  = \mathcal{N}_{\check{F}^{\sigmab}_1}^{(1)}\,,\nn\\
  &\frac{1}{\Rc^2}\nabla_{\psib}\Phi_{\psi}
  -\frac{\tau e^\varphi}{2\Rc}\sD^A\Phi_A
  -\frac{p}{\Rc^2}(\mathcal{H}-\Phi_{\psib})
  =\mathcal{N}_{\check{F}^{\sigmab}_1}^{(2)}\,,\nn\\
  &\tau\nabla_T\Phi_A+W^{(3)}_\phi -\cpr\sD_A\Phi_{\psib}
  + \frac{\cmr}{\Rc}\sD_A\Phi_{\psi}
  = \mathcal{N}_\phi^{(3)}\Rc\,,\nn\\
  &\nabla_{\psib}\Phi -\Phi_{\psib}=0\,,
\end{align}
where the stratified null forms~\eqref{SNFs1} still apply.  Following
the same procedure that resulted in~\eqref{EqsRescaled} we obtain for
the inverse areal radius~$\Rc^{-1}$,
\begin{align}\label{EqsRescaledRc}
  &\frac{1}{\Rc^2}\nabla_\psi\rho_{\psib}
  - \frac{\tau e^\varphi}{2\Rc^2}\sD^A\rho_A
  =-\tilde{\mathcal{N}}_{\Rc^{-1}}
  -\frac{\tau e^\varphi}{\Rc^3}\rho_A\rho^A\,,\nn\\
  &\frac{1}{\Rc^3}\nabla_{\psib}\rho_\psi
  -\frac{\tau e^\varphi}{2\Rc^2}\sD^A\rho_A
  = \mathcal{N}_{\Rc^{-1}}^{(2)}\,,\nn\\
  &\tau\nabla_T\rho_A+W^{(3)}_\rho-\cpr\sD_A\rho_{\psib}
  +\frac{\cmr}{\Rc}\sD_A\rho_{\psi}
  = \mathcal{N}_{\Rc^{-1}}^{(3)}\Rc^2\,,\nn\\
  &\nabla_{\psib}\Rc^{-1} =\frac{1}{\Rc^2}(1-\rho_{\psib})\,,
\end{align}
where the stratified null forms on the RHSs are defined as,
\begin{align}
  &\mathcal{N}_{\Rc^{-1}}^{(2)}:= -\tilde{\mathcal{N}}_{\Rc^{-1}}
  - \frac{\rho_\psi(1-\rho_{\psib})}{\Rc^4}
  - \frac{\tau e^\varphi\rho_A\rho^A}{\Rc^3}\nn\\
  &\quad\quad +\frac{1}{\tau\Rc^2}\left(\mathcal{C}_{+\psib}^R
  -\mathcal{C}_{-\psi}^R
  -\frac{\tau e^\varphi}{2}Z^\sigma\right)\left(2-\rho_{\psib}
  +\frac{\rho_{\psi}}{\Rc}\right)\nn\,,\\
  &\mathcal{N}_{\Rc^{-1}}^{(3)}=\frac{\cmr}{\Rc^4}\rho_\psi\rho_A\\
  &+\left(\frac{1}{\Rc^2}+\frac{\rho_\psi}{\Rc^3}\right)
  \left(-\frac{\cpr}{\tau}\tau_{,A}
  -\frac{\mathcal{C}_A}{\tau}\nabla_T\cpr
  +\nabla_{T}\mathcal{C}^-_A-\sD_A\cmr\right)\nn\\
  &-\left(\frac{1}{\Rc^2}-\frac{\rho_{\psib}}{\Rc^2}\right)
  \left(\frac{\cmr}{\tau}\tau_{,A}
  +\frac{\mathcal{C}_A}{\tau}\nabla_T\cmr
  +\nabla_{T}\mathcal{C}^+_A+\sD_A\cpr\right)\,.\nn
\end{align}
We now have all the wave equations in first order form with all
variables rescaled.

\paragraph*{Rescaled reduction constraints:} With the exception
of~$\Rc^{-1}$ all variables must satisfy the following reduction
constraints,
\begin{align}
  \Rc\nabla_\psi\Phi -\Phi_{\psi} =0\,,\quad\sD_A\Phi-\Phi_A=0\,,
\end{align}
whereas~$\Rc^{-1}$ must satisfy,
\begin{align}
  &\nabla_\psi(\Rc^{-1})+\frac{1}{\Rc^2}
  \left(1+\frac{\rho_\psi}{\Rc}\right)=0\,,\nn\\
  &\sD_A(\Rc^{-1})+\frac{\rho_A}{\Rc^2}=0\,,
\end{align}
This concludes the rescaling of the variables and we are now ready to
compactify the radial coordinate and move to hyperboloidal slices.

\section{Hyperboloidal Compactification}
\label{section:compactification}

In order to write the evolution equations in their final form, we want
to define a radially compactified hyperboloidal coordinate system in
much the same way as was done in~\cite{GasGauHil19}. However, we want
to work with the areal radius~$\Rc$ as our preferred radial coordinate
rather than~$R$. So as an intermediate step we do another coordinate
change.

\subsection{From radius to areal radius}

Let us consider the coordinate
system~$(\mathring{T},\Rc,\mathring{\theta}^A)$ related
to~$(T,R,\theta^A)$ by,
\begin{align}\label{2ndCoords}
  \mathring{T}=T,\quad \Rc= R e^{\epsilon/2},
  \quad\mathring{\theta}^A=\theta^A\,,
\end{align}
and let us define the outgoing and incoming null
vectors~$\mathring{\psi}$ and~$\mathring{\psib}$ in a way analogous
to~$\psi$ and~$\psib$,
\begin{align}\label{psiring}
  \mathring{\psi}^a&=\p_{\mathring{T}}^a
  +\mathcal{C}^{\Rc}_+\p_{\Rc}^a\,,\quad
  \mathring{\psib}^a=\p_{\mathring{T}}^a
  +\mathcal{C}^{\Rc}_-\p_{\Rc}^a\,.
\end{align}
We define the null covectors~$\mathring{\sigma}$
and~$\mathring{\sigmab}$ analogously to~$\sigma$
and~$\sigmab$. Similarly we define the
quantities~$\mathring{\varphi}$, $\mathring{\tau}$,
$\mathring{\sg}^{ab}$ and~$\mathring{\mathcal{C}}_A^\pm$. We can write
the vectors~$\p_T^a$ and~$\p_R^a$ in terms of~$\p_{\mathring{T}}^a$
and~$\p_{\mathring{R}}^a$ in the following way,
\begin{align}\label{pTpRTopT'pR'}
  &\p_T^a = \p_{\mathring{T}}^a
  + \frac{1}{\tau}
  \left(\cpr\nabla_{\psib}\Rc-\cmr\nabla_{\psi}\Rc\right)
  \p_{\mathring{R}}^a\,,\nn\\
  &\p_R^a = \frac{1}{\tau}
  \left(\nabla_{\psi}\Rc-\nabla_{\psib}\Rc\right)
  \p_{\mathring{R}}^a\,,
\end{align}
and from~\eqref{pTpRTopT'pR'} a straightforward calculation leads to,
\begin{align}\label{psiRcTopsi}
  \mathring{\psi}^a&=\frac{1}{\nabla_\psi\Rc-\nabla_{\psib}\Rc}
  \left[\left(\mathcal{C}^{\Rc}_+-\nabla_{\psib}\Rc \right)\psi^a
    +\left(\nabla_{\psi}\Rc-\mathcal{C}^{\Rc}_+ \right)\psib^a\right]\,,\nn\\
  \mathring{\psib}^a&=
  \frac{1}{\nabla_\psi\Rc-\nabla_{\psib}\Rc}\left[\left(\mathcal{C}^{\Rc}_-
    -\nabla_{\psib}\Rc \right)\psi^a
    +\left(\nabla_{\psi}\Rc-\mathcal{C}^{\Rc}_- \right)\psib^a\right]\,.
\end{align}
At this point, in order to complete the expressions that relate the
null vectors that correspond to either coordinate system, we need only
to find an expression for the coordinate light
speeds~$\mathcal{C}^{\Rc}_+$ and~$\mathcal{C}^{\Rc}_-$. This is given
by,
\begin{align}\label{CpRcToCpR}
  &\mathcal{C}^{\Rc}_+ = \frac{J^{\Rc}{}_T
    +\cpr J^{\Rc}{}_R}{J^{\mathring{T}}{}_T+\cpr J^{\mathring{T}}{}_R}
  = \nabla_\psi\Rc\,,\nn\\
  &\mathcal{C}^{\Rc}_- = \frac{J^{\Rc}{}_T
    +\cmr J^{\Rc}{}_R}{J^{\mathring{T}}{}_T+\cmr J^{\mathring{T}}{}_R}
  = \nabla_{\psib}\Rc\,.
\end{align}
Plugging~\eqref{CpRcToCpR} into~\eqref{psiRcTopsi} we find that,
\begin{align}\label{psipsic}
  \mathring{\psi}^a = \psi^a\,,\quad \mathring{\psib}^a
  = \psib^a\,.
\end{align}
In other words, the expression for our incoming and outgoing null
vectors are invariant under the coordinate
change~$(T,R,\theta^A)\rightarrow(\mathring{T},\Rc,\mathring{\theta}^A)$.

\paragraph*{Angular derivatives~$\sD_a$:} With~\eqref{psipsic} we
already know how null derivatives transform under this coordinate
change, so we are only lacking an expression that relates~$\sD_a$
to~$\mathring{\sD}_a$, where the latter is defined analogously to the
former. We contract~$\mathring{\sigma}$ with~$\p_{\Rc}$ in order to
find the relation between~$\varphi$ and~$\mathring{\varphi}$,
\begin{align}
	e^{\varphi-\mathring{\varphi}}=\frac{\mathring{\tau}}{\tau}\,.
\end{align}
Under this change of coordinates the angular basis vectors transform
in the following way,
\begin{align}
	\p_{\theta^A}^a=(\p_{\theta^A}\Rc)\p_{\Rc}^a+\p_{\mathring{\theta}^A}^a\,,
\end{align}
and we can use this to find the relation between~$\mathcal{C}_A^{\pm}$
and~$\mathring{\mathcal{C}}_A^{\pm}$. By recognizing that,
\begin{align}
  \mathring{\mathcal{C}}_A^{\pm} = \mathring{\sigma}_a\p_{\mathring{\theta}^A}^a\,,
\end{align}
we find that,
\begin{align}\label{CAcCA}
  &\mathring{\mathcal{C}}_A^+
  =-\sD_A\Rc+\frac{1}{\tau}(\mathcal{C}^-_A\nabla_{\psib}\Rc
  +2\mathcal{C}^+_A\nabla_\psi\Rc-\mathcal{C}^+_A\nabla_{\psib}\Rc)\,,\nn\\
  &\mathring{\mathcal{C}}_A^-
  =\sD_A\Rc-\frac{1}{\tau}(2\mathcal{C}^-_A\nabla_{\psib}\Rc
  +\mathcal{C}^+_A\nabla_\psi\Rc-\mathcal{C}^-_A\nabla_{\psi}\Rc)\,.
\end{align}
We can now build the derivatives by expanding the components
of~$\mathring{\sg}_a{}^b$ in mixed form,
\begin{align}
  \mathring{\sD}_A\phi = \mathring{\sg}_A{}^{\mathring{T}}\p_{\mathring{T}}\phi
  +\mathring{\sg}_A{}^{\Rc}\p_{\Rc}\phi
  +\mathring{\sg}_A{}^{B}\p_{\mathring{\theta}^B}\phi\,.
\end{align}
Using~\eqref{sgMixed} the non-trivial components
of~$\mathring{\sg}_a{}^b$ can be written as,
\begin{align}
  &\mathring{\sg}_A{}^{\mathring{T}}
  = \frac{\mathcal{C}_A}{\tau}\,,\quad\mathring{\sg}_A{}^{\mathring{R}}
  =\frac{\mathcal{C}_A^+\nabla_\psi\Rc+\mathcal{C}_A^-\nabla_{\psib}\Rc}{\tau}\,.
\end{align}
A straightforward calculation then leads to,
\begin{align}
  \mathring{\sD}_A\phi =\sD_A\phi\,,
\end{align}
which shows that under this coordinate transformation, all derivatives
stay the same.

\subsection{Hyperboloidal Compactification} 

Let us consider a third coordinate system~$(t,r,\bar{\theta}^A)$ which
is related to the previous one by,
\begin{align}\label{3rdCoords}
  \mathring{T}=t+H(\Rc(r),\mathring{\theta}^A),\quad \Rc
  = \Rc(r),\quad\bar{\theta}^A=\mathring{\theta}^A\,,
\end{align}
where~$H(\Rc(r),\mathring{\theta}^A)$ and~$\Rc( r)$ are called height
and compression functions, respectively. Note that this coordinate
change differs from the one in~\cite{GasGauHil19} insofar as the
height function depends on the angular coordinates. Once again we
define a set of null vectors associated to~\eqref{3rdCoords},
\begin{align}
  \xi^a&=\p_{ t}^a+\mathcal{C}^{ r}_+\p_{ r}^a\,,\quad
  \ul{\xi}^a=\p_{ t}^a+\mathcal{C}^{ r}_-\p_{ r}^a\,.
\end{align}
For this set of coordinates we define~$\bar{\sigma}$, $\bar{\sigmab}$,
$\bar{\varphi}$, $\bar{\tau}$, $\bar{\sg}^{ab}$
and~$\bar{\mathcal{C}}_A^\pm$.The relations between the first two
coordinate basis vectors can be written as,
\begin{align}
  \p_{ t}^a = \p_{\mathring{T}}^a\,,\quad\p_{ r}^a
  = H'\Rc'\p_{\mathring{T}}^a+\Rc'\p_{\mathring{R}}^a\,,
\end{align}
where~$H':=\p_{\Rc}H$ and $\Rc':=\p_{r}\Rc$, which yields,
\begin{align}\label{xiToTR}
  &\xi^a = (1+H'\Rc'\mathcal{C}^{\Rc}_+)\p_{\mathring{T}}^a
  + \Rc'\mathcal{C}^{\Rc}_+\p_{\Rc}^a\,,\nn\\
  &\ul{\xi}^a = (1+H'\Rc'\mathcal{C}^{\Rc}_-)\p_{\mathring{T}}^a
  + \Rc'\mathcal{C}^{\Rc}_-\p_{\Rc}^a\,.
\end{align}
The last ingredient we need is the expressions that relate the
coordinate light speeds. We find this for~$\mathcal{C}^{ r}_+$ as the
derivation of~$\mathcal{C}^{ r}_-$ is completely analogous,
\begin{align}\label{CpRcToCpr}
  \mathcal{C}^{\Rc}_+ &= \frac{J^{\Rc}{}_t
    +\cpr J^{\Rc}{}_r}{J^{\mathring{T}}{}_t+\cpr J^{\mathring{T}}{}_r}
  = \frac{\nabla_\xi\Rc}{\nabla_\xi\mathring{T}}\nn\\
  &= \frac{\mathcal{C}^{ r}_+\Rc'}{1+\mathcal{C}^{ r}_+H'\Rc'}\,.
\end{align}
Inverting~\eqref{CpRcToCpr} we get,
\begin{align}\label{CpRcToCpr2}
  \mathcal{C}^{ r}_\pm =
  \frac{\mathcal{C}^{\Rc}_\pm}{\Rc'(1-\mathcal{C}^{\Rc}_\pm H')}\,,
\end{align}
where we have also included the expression
for~$\mathcal{C}^{r}_-$. Substituting~\eqref{CpRcToCpr2}
into~\eqref{xiToTR} and using~\eqref{psiring} we get,
\begin{align}
  &\xi^a = \frac{1}{1-H'\mathcal{C}^{\Rc}_+}\mathring{\psi}^a\,,
  \quad
  \ul{\xi}^a = \frac{1}{1-H'\mathcal{C}^{\Rc}_-}\mathring{\psib}^a\,.
\end{align}
Finally, using the invariance of the null vectors under the first
coordinate change we find the relation between~$\psi^a$ and~$\psib^a$,
and~$\xi^a$ and~$\ul{\xi}^a$,
\begin{align}\label{psiToxi}
  \psi^a =\Omega^+\xi^a \,,\quad\psib^a = \Omega^-\ul{\xi}^a\,,
\end{align}
where the scalar factors~$\Omega^\pm$ are defined as,
\begin{align}
  \Omega^+:=1-H'\nabla_\psi\Rc \,,\quad
  \Omega^-:=1-H'\nabla_{\psib}\Rc\,.
\end{align}

\paragraph*{Angular derivatives~$\sD_a$:} In order to finish the
compactification of the coordinates we only need to find
how~$\mathring{\sD}_a$ transforms. In other words, we want to find a
relation between~$\bar{\sD}_a$ and~$\sD_a$, where the former is
defined in the obvious way. The steps of the procedure are the same,
so we present only the relevant expressions in this derivation,
\begin{align}
  &e^{\mathring{\varphi}-\bar{\varphi}}=\frac{1}{\Rc'}\,,
  \quad\p_{\bar{\theta}^A}^a=(\p_{\mathring{\theta}^A}H)
  \p_{\mathring{T}}^a + \p_{\mathring{\theta}^A}^a\,,\\
  &\bar{\mathcal{C}}_A^+=\frac{\mathring{\mathcal{C}}_A^+
    -\nabla_\psi\Rc(\p_{\mathring{\theta}^A}H)}{\Rc'\Omega^+}\,,
  \quad
  \bar{\mathcal{C}}_A^-=\frac{\mathring{\mathcal{C}}_A^-
    +\nabla_{\psib}\Rc(\p_{\mathring{\theta}^A}H)}{\Rc'\Omega^-}\,.\nn
\end{align}
A rather lengthy calculation leads to the following result,
\begin{align}\label{sDTosDbar}
  \sD_A\phi =& \bar{\sD}_A\phi - H'\left[\mathring{\mathcal{C}}_A^+
    \nabla_\xi\phi-\mathring{\mathcal{C}}_A^-
    \nabla_{\ul{\xi}}\phi\right]\,,
\end{align}
where~$\mathring{\mathcal{C}}_A^\pm$ are given in terms of our metric
variables by~\eqref{CAcCA}.

\section{Compactified EFEs in first order form}
\label{section:EFEsFirstOrder}

We are now equipped to present the EFEs regularized at null infinity
in first order form and in radially compactified coordinates. The
equations for the metric functions~$\mathcal{C}_\pm^R\mp
1$,~$\mathcal{\hat{C}}^\pm_A$, $\varphi$, $h_+$ and~$h_\times$ are
\begin{align}\label{finalfinal1}
  &\frac{\Omega^+}{\Rc^{q+1}}\nabla_\xi\left[\Rc^q\Phi_{\psib}\right]
  -\frac{\tau e^\varphi}{2\Rc}\sD^A\Phi_A = \mathcal{N}_\phi^{(1)}\,,\nn\\
  &\frac{\Omega^-}{\Rc^2}\nabla_{\ul{\xi}}\Phi_{\psi}
  -\frac{\tau e^\varphi}{2\Rc}\sD^A\Phi_A+\frac{q}{\Rc^2}\Phi_{\psib}
  =\mathcal{N}_\phi^{(2)}\,,\nn\\
  &\cpr\Omega^-\nabla_{\ul{\xi}}\Phi_A -\cmr\Omega^+\nabla_{\xi}\Phi_A
  -\cpr\sD_A\Phi_{\psib} + \frac{\cmr}{\Rc}\sD_A\Phi_{\psi} \nn\\
  &\quad\quad\quad\quad\quad\quad\quad\quad\quad\quad\quad\quad+W^{(3)}_\phi
  = \Rc\mathcal{N}_\phi^{(3)},\nn\\
  &\Omega^-\nabla_{\ul{\xi}}\Phi -\Phi_{\psib}=0\,,
\end{align}
where~$p$ is set to~$0$ in the cases of~$h_+$ and~$h_\times$ and the
we use~$\sD^A$ merely as a shorthand for the expression on the RHS
of~\eqref{sDTosDbar} with raised indices. The equations
for~$\check{F}^{\sigmab}_1$ are,
\begin{align}\label{finalfinal2}
  &\frac{\Omega^+}{\Rc^{p+1}}\nabla_\xi\left[\Rc^p\Phi_{\psib}\right]
  -\frac{\tau e^\varphi}{2\Rc}\sD^A\Phi_A-\frac{p}{\Rc^2}\mathcal{H}
  = \mathcal{N}_{\check{F}^{\sigmab}_1}^{(1)}\,,\nn\\
  &\frac{\Omega^-}{\Rc^2}\nabla_{\ul{\xi}}\Phi_{\psi}
  -\frac{\tau e^\varphi}{2\Rc}\sD^A\Phi_A-\frac{p}{\Rc^2}
  (\mathcal{H}- \Phi_{\psib})
  =\mathcal{N}_{\check{F}^{\sigmab}_1}^{(2)}\,,\nn\\
  &\cpr\Omega^-\nabla_{\ul{\xi}}\Phi_A
  -\cmr\Omega^+\nabla_{\xi}\Phi_A  -\cpr\sD_A\Phi_{\psib}
  + \frac{\cmr}{\Rc}\sD_A\Phi_{\psi} \nn\\
  &\quad\quad\quad\quad\quad\quad\quad\quad\quad\quad\quad\quad
  +W^{(3)}_\phi= \Rc\mathcal{N}_\phi^{(3)},\nn\\
  &\Omega^-\nabla_{\ul{\xi}}\Phi -\Phi_{\psib}=0\,,
\end{align}
and the ones for~$\Rc^{-1}$ are,
\begin{align}\label{finalfinal3}
  &\frac{\Omega^+}{\Rc^2}\nabla_\xi\rho_{\psib}
  - \frac{\tau e^\varphi}{2\Rc^2}\sD^A\rho_A
  =-\tilde{\mathcal{N}}_{\Rc^{-1}}
  -\frac{\tau e^\varphi}{\Rc^3}\rho_A\rho^A\,,\nn\\
  &\frac{\Omega^-}{\Rc^3}\nabla_{\ul{\xi}}\rho_\psi
  -\frac{\tau e^\varphi}{2\Rc^2}\sD^A\rho_A
  = \mathcal{N}_{\Rc^{-1}}^{(2)}\,,\nn\\
  &\cpr\Omega^-\nabla_{\ul{\xi}}\rho_A-\cmr\Omega^+\nabla_{\xi}\rho_A
  -\cpr\sD_A\rho_{\psib}+\frac{\cmr}{\Rc}\sD_A\rho_{\psi}\nn\\
  &\quad\quad\quad\quad\quad\quad\quad\quad\quad\quad\quad\quad
  +W^{(3)}_\rho=\Rc^2\mathcal{N}_{\Rc^{-1}}^{(3)},\nn\\
  &\Omega^-\nabla_{\ul{\xi}}\Rc^{-1}
  =\frac{1}{\Rc^2}(1-\rho_{\psib})\,,
\end{align}

\paragraph*{Compactified reduction constraints:} The final form of the
reduction constraints for all variables except~$\Rc^{-1}$ is,
\begin{align}\label{finalConstraints1}
  \Rc\Omega^+\nabla_\xi\Phi -\Phi_{\psi}
  =0\,,\quad\sD_A\Phi-\Phi_A=0\,,
\end{align}
whereas~$\Rc^{-1}$ must satisfy,
\begin{align}\label{finalConstraints2}
  &\Omega^+\nabla_\xi(\Rc^{-1})+\frac{1}{\Rc^2}
  \left(1+\frac{\rho_\psi}{\Rc}\right)=0\,,\nn\\
  &\sD_A(\Rc^{-1})+\frac{\rho_A}{\Rc^2}=0\,,
\end{align}
We now have the complete set of equations written in a radially
compactified coordinate
system. Equations~\eqref{finalfinal1},~\eqref{finalfinal2}
and~\eqref{finalfinal3} are the EFEs in GHG written as a system of
first order differential equations that are regular at null infinity,
with reduction constraints given by~\eqref{finalConstraints1}
and~\eqref{finalConstraints2}.

\subsection{Comment on the choice of reduction variables}

Thus far we have written the first order equations in the third
coordinate system~\eqref{3rdCoords}, which we referred to as radially
compactified coordinates. We have written the derivatives in the EFEs
in terms of the compression and height functions~$\Rc(r)$
and~$H(\Rc(r),\mathring{\theta}^A)$, but we have said nothing so far
about how we want these functions to behave. Strictly speaking, we
have not done a \textit{compactification} yet. To do that we assume,
\begin{align}\label{RprimeAssump}
	\Rc'(r)\simeq \Rc^n\,,
\end{align}
where~$1<n\leq 2$. The lower bound on~$n$ is necessary to make~$r$
approach a finite value as~$\Rc$ goes to null infinity,
whereas~$0<n\leq 2$ is required for numerical stability, as discussed
in~\cite{CalGunHil05}. Moreover, we want~$\mathcal{C}^r_+\simeq 1$
which, together with~\eqref{RprimeAssump} gives us the requirement on
the leading order behavior of~$H$,
\begin{align}\label{leadingH}
  H'(\Rc(r),\mathring{\theta}^A)\simeq 1-\frac{m_{\cpr,1}}{\Rc}
  -\frac{1}{\Rc^{n}}\,,
\end{align}
where~$m_{\cpr,1}$ is the leading order term of the
field~$\Rc(\cpr-1)$ and depends only on the angular coordinates. Our
requirements on the functional dependence of~$H$ forbid the inclusion
of higher order time dependent corrections, but could include higher
order corrections in~$m_{\cpr,1}$. Fortunately, provided that~$n<2$
the condition~$\mathcal{C}^r_+\simeq1$ is satisfied
with~\eqref{leadingH} anyway. In the desirable~$n=2$ case we instead
obtain~$\mathcal{C}^r_+=O(1)$, provided that~$p\geq2$ so that no logs
are present to order~$\Rc^{-2}$, which is acceptable also. For
all~$1<n\leq2$ we end up with a formally singular term
in~$\mathcal{C}^r_+$ that is easily evaluated with L'H\^opital's
rule. To obtain instead the sharper condition~$\mathcal{C}^r_+=1$ over
the full range of~$n$ we could instead solve the eikonal equation as
proposed in~\cite{HilHarBug16}. Plugging this in~\eqref{psiToxi} we
get that the outgoing and incoming null vectors transform
asymptotically as,
\begin{align}
  \Omega^+ \simeq \frac{1}{\Rc^n}\,,\quad \Omega^-\simeq O(1)\,.
\end{align}
This has a very significant influence on how far we can go in
rescaling reduction variables, because replacing~$\psi^a$ with~$\xi^a$
gives an extra~$n$ powers of~$\Rc^{-1}$, whereas~$\psib^a$ does
not. Take the first equation in~\eqref{finalfinal1}, for example. As
the stratified null forms are~$o^+(R^{-2})$, we can write
asymptotically,
\begin{align}\label{whyrescalePhipsib}
  &\frac{1}{\Rc^{q+1+n}}\nabla_\xi\left[\Rc^q\Phi_{\psib}\right]
  =o^+(R^{-2})\,,\nn\\
  \Rightarrow & \frac{1}{\Rc^{1+n}}\nabla_\xi\Phi_{\psib}
  -q(\Rc^{-1})_{,\psi}\Phi_{\psib}=o^+(R^{-2})\,,\nn\\
  \Rightarrow & \nabla_\xi\Phi_{\psib}
  -q\Rc^{1+n}(\Rc^{-1})_{,\psi}\Phi_{\psib}=o^+(R^{n-1})\,.
\end{align}
According to the findings in~\cite{DuaFenGasHil21}, the only
possibility for the error terms on the RHS of the second line to decay
slower than~$O(R^{n-2})$ is if there are terms proportional to~$\log
R$. As we know that such terms are suppressed up to order~$q$, if this
number is sufficiently large we can write that,
\begin{align}
  \nabla_\xi\Phi_{\psib}+\frac{q\Phi_{\psib}}{\Rc}=O(R^{n-2})\,.
\end{align}
This guarantees that when we integrate numerically along integral
curves of~$\xi^a$, the integral will not act upon terms that diverge
at null infinity. If we had rescaled~$\Phi_{\psib}$ by another power
of~$\Rc$, then we would have had to integrate an error term of the
type~$O(R^{n-1})$ which always diverges. This is the reason why we
cannot afford to have this particular reduction variable approach a
(non-vanishing) finite value at null infinity. It is worth pausing
here for a moment to analyze what~\eqref{leadingH} means.
\begin{remark}
  Our hyperboloidal compactification is different from others done in
  the literature, for example in~\cite{Hil16} and~\cite{GauVanHil21},
  in the sense that we have allowed the height function~$H$ to depend
  upon the angular coordinates~\eqref{3rdCoords}. The reason for this
  is that~$H'$ depends on~$m_{\cpr,1}$ and we know
  from~\cite{DuaFenGas22} that that function, the numerator at first
  order in~$\Rc^{-1}$ in the polyhomogeneous expansion of~$\cpr$, is
  an angular function, as it would be for a general ugly field in a
  curved spacetime. If~$\cpr$ were a good, then~$m_{\cpr,1}$ would be
  a radiation field, meaning that it would be allowed to vary along
  integral curves of~$\psib^a$. As a consequence,~$H$ would have to
  vary with~$T$ as well as all other coordinates and this type of
  compactification~\eqref{3rdCoords} would not work.
\end{remark}

\subsection{Formally singular terms}

When implementing this system numerically, it is helpful if the terms
we integrate have explicitly regular limits at null infinity, as
opposed to terms which can only be written as the quotient of
divergent terms, for instance~$O(R)/O(R)$. These \textit{formally
  singular terms} are implicitly regular as they acquire regular
values with the asymptotics we expect for each field, but they cause
problems in the implementation nonetheless. They need to be identified
and carefully processed using L'H\^opital's rule before the numerical
implementation, as was done for example in~\cite{MonRin08}. This
subsection is dedicated to identifying the terms
in~\eqref{finalfinal1},~\eqref{finalfinal2} and~\eqref{finalfinal3}
that contribute to subleading order and are formally singular so they
can later be treated separately, as well as to choosing the constraint
addition~$W^{(2)}_{ab}$ so that the null forms~$\mathcal{N}_\phi$
cannot possibly have these terms.

\paragraph*{$\mathcal{N}^{(i)}_\phi$ in terms of $\mathcal{N}_\phi$:}
For the sake of tidiness, a lot of terms in every equation have been
put together into groups of SNFs throughout this work to allow us to
focus on the principal terms and those that contribute to leading
order. However, because the goal of this paper is ultimately to write
down a set of equations that are ready to be numerically implemented,
we have to be able write all terms in all equations explicitly as
products of derivatives of metric functions. In order to help the
reader navigate through these SNFs we gather those terms in this
paragraph. The SNFs named as~$\mathcal{N}^{(i)}_\phi$ are defined as,
\begin{align}\label{SNFs1Tidy}
	& \mathcal{N}_\phi^{(1)} :=\tilde{\mathcal{N}}_\phi-
  \Phi\tilde{\mathcal{N}}_{\Rc^{-1}}+\frac{\Phi_\psi}{\Rc^3}
  (\rho_{\psib}-1) - \frac{\tau e^\varphi\Phi^A}{\Rc^2}
  \rho_A\,,\nn\\ & \mathcal{N}_\phi^{(2)}:=
  \frac{1}{\tau\Rc}\left(\mathcal{C}_{+,\psib}^R-\mathcal{C}_{-,\psi}^R
  -\frac{\tau
    e^\varphi}{2}Z^\sigma\right)\left(\frac{\Phi_{\psi}}{\Rc}
  -\Phi_{\psib}\right)
  \nn\\ -&\frac{q}{\Rc^3}\Phi_{\psib}\rho_\psi+
  \frac{2\Phi_\psi}{\Rc^3}(\rho_{\psib}-1)- \frac{\tau
    e^\varphi\Phi^A}{\Rc^2} \rho_A+\tilde{\mathcal{N}}_\phi
  -\Phi\tilde{\mathcal{N}}_{\Rc^{-1}}\,,\nn\\ &
  \mathcal{N}_\phi^{(3)}
  =\frac{1}{\Rc}\Phi_{\psib}\left(-\frac{\cpr}{\tau}\tau_{,A}
  -\frac{\mathcal{C}_A}{\tau}\nabla_T\cpr+\nabla_{T}\mathcal{C}^-_A
  -\sD_A\cmr\right)\nn\\
  +&\frac{1}{\Rc^2}\Phi_{\psi}\left(2\cmr\rho_A+\frac{\cmr}{\tau}\tau_{,A}
  +\frac{\mathcal{C}_A}{\tau}\nabla_T\cmr+\nabla_{T}\mathcal{C}^+_A
  +\sD_A\cpr\right),
\end{align}
for all variables except~$\Rc^{-1}$, whereas for the latter they are,
\begin{align}
  &\mathcal{N}_{\Rc^{-1}}^{(2)}:= -\tilde{\mathcal{N}}_{\Rc^{-1}}
  - \frac{\rho_\psi(1-\rho_{\psib})}{\Rc^4}
  - \frac{\tau e^\varphi\rho_A\rho^A}{\Rc^3}\nn\\
  &\quad\quad +\frac{1}{\tau\Rc^2}\left(\mathcal{C}_{+\psib}^R
  -\mathcal{C}_{-\psi}^R
  -\frac{\tau e^\varphi}{2}Z^\sigma\right)\left(2-\rho_{\psib}
  +\frac{\rho_{\psi}}{\Rc}\right)\nn\,,\\
  &\mathcal{N}_{\Rc^{-1}}^{(3)}=\frac{\cmr}{\Rc^4}\rho_\psi\rho_A\\
  &+\left(\frac{1}{\Rc^2}+\frac{\rho_\psi}{\Rc^3}\right)
  \left(-\frac{\cpr}{\tau}\tau_{,A}-\frac{\mathcal{C}_A}{\tau}
  \nabla_T\cpr+\nabla_{T}\mathcal{C}^-_A-\sD_A\cmr\right)\nn\\
  &-\left(\frac{1}{\Rc^2}-\frac{\rho_{\psib}}{\Rc^2}\right)
  \left(\frac{\cmr}{\tau}\tau_{,A}
  +\frac{\mathcal{C}_A}{\tau}\nabla_T\cmr
  +\nabla_{T}\mathcal{C}^+_A+\sD_A\cpr\right)\,.\nn
\end{align}
The SNFs named~$\tilde{\mathcal{N}}_\phi$ are defined in terms
of~$\mathcal{N}_\phi$ as,
\begin{align}
  \tilde{\mathcal{N}}_\phi = &- \frac{\tau e^\varphi}{2}\mathcal{N}_\phi
  +\frac{e^\varphi}{2}\left(\frac{\mathcal{C}_A}{\tau}\sD^A\cpr
  -\sD^A\mathcal{C}^+_A\right)\nabla_{\psib}\phi\nn \\
  &- \frac{e^\varphi}{2}\left(\frac{\mathcal{C}_A}{\tau}\sD^A\cmr
  +\sD^A\mathcal{C}^-_A\right)\nabla_{\psi}\phi\nn\\
  &+ \frac{1}{\tau}\nabla_\psi\cmr(\nabla_\psi \phi-\nabla_{\psib}\phi)\,.
\end{align}
Finally, the SNFs~$\mathcal{N}_\phi$ for each of the variables are
exactly the RHSs of~\eqref{finaleqs}.

\paragraph*{Identifying formally singular terms:} In order to single
out which terms might cause problems, one needs to look at each
equation carefully and multiply it through by whatever required power
of~$\Rc$ is in front of the leading second order term. For example,
in~\eqref{finalfinal1}, the last two equations have no~$\Rc$ in front
of them, so nothing needs to be done. The second equation needs to be
multiplied through by~$\Rc^3$, whereas the first one should be
multiplied by~$\Rc^{1+n}$, as can be seen
in~\eqref{whyrescalePhipsib}. After that, the terms we are looking for
will be those that still have some explicit positive power of~$\Rc$
left over and contribute to subleading order. We begin
with~\eqref{finalfinal1}. The first equation has a formally singular
term on the LHS upon expanding the~$\nabla_\xi$ derivative
\begin{align}\label{formallysing1}
  q\Rc^{n-1}\Phi_{\psib}\,.
\end{align} 
On the RHS there are no such terms outside~$\mathcal{N}_\phi$ and we
leave the analysis of~$\mathcal{N}_\phi$ to the next paragraph. Bear
in mind that upstairs~$A$ indices have an implicit~$\sg^{ab}$, which
contains an explicit~$\Rc^{-2}$,
see~\eqref{eqn:qconformaltransformation}. The second equation
in~\eqref{finalfinal1} contains no formally singular terms on the LHS
and none on the RHS outside~$\mathcal{N}_\phi$. The third and fourth
equations have no formally singular terms at all. Note that in the
good equations, namely~$h_+$ and~$h_\times$, terms
like~\eqref{formallysing1} vanish since~$q=0$. In~\eqref{finalfinal2}
we find similar terms. In the first equation we have,
\begin{align}\label{formallysing3}
	p\Rc^{n-1}(\Phi_{\psib}-H)\,.
\end{align}
Note that the reduction variable associated with a bad derivative of
the gauge driver does not necessarily have decay, since the gauge
driver is not an ugly, so this term seems like it diverges. However,
we have built~$\check{F}_1^{\sigmab}$ in such a way
that~$(\Phi_{\psib}-H)$ decays~\eqref{f-hdecay}. Therefore these terms
are only formally singular. In~\eqref{finalfinal3} there are no
formally singular terms outside~$\mathcal{N}_{\Rc^{-1}}$. The only
thing left to do is then to check if any of these terms show up in the
original stratified null forms.

\paragraph*{Formally singular terms in~$\mathcal{N}_\phi$:} With the
help of computer algebra we typed the wave equations in second order
form~\eqref{finaleqs} and substituted the gauge source functions and
constraint additions with the choices that suppress logs up to
order~$p$,~\eqref{gaugefinal} and~\eqref{eqn:ConstAdd2}. Once the EFEs
were written in terms of~$(\Gamma\Gamma)_{ab}$ functions, we replaced
them with derivatives of the metric functions using all the components
of~$\Gb_a{}^b{}_c$. Then we rescaled all the variables and their
derivatives in order to make the powers of~$\Rc^{-1}$ explicit and
most of the equations have formally singular terms that arise from the
fact that bad derivatives of ugly fields have an
implicit~$\Rc^{-1}$. So terms that are~$O(R^{-3})$ but contain a bad
derivative of an ugly are necessarily formally singular. However, all
these bad derivatives act upon either~$\cpr$, $\Rc^{-1}$
or~$\mathcal{C}_A^+$, precisely the fields associated with the~$4$
constraints. Because the constraints are essentially bad derivatives
of these functions to leading order, we can add specific combinations
of the former so the latter do not appear at third order. This is the
reason why we kept the subleading constraint addition~$W^{(2)}_{ab}$
free in the equations for the~$10$ metric variables. The constraints
we have to add to each of the equations so that the stratified null
forms~$\mathcal{N}_\phi$ contain no formally singular terms are,
\begin{align}
  &W^{(2)}_{\psib\psib} = \frac{\nabla_{\psib}\Rc}{\Rc}Z^{\sigmab}
  +\frac{\nabla_{\psib}\cmr}{\tau}Z^\sigma\,,\nn\\
  &W^{(2)}_{(\psi\psib)} = \frac{1}{\tau\Rc}Z^{\sigmab}
  +\frac{\nabla_{\psib}\varphi}{\tau}Z^\sigma
  +\frac{\nabla_{\psib}\mathcal{C^+_A}}{\tau}Z^A\,,\nn\\
  &W^{(2)}_{\psi A} = -\frac{\nabla_{\psib}\mathcal{C}_A^-}{\tau}Z^\sigma\,,\nn\\
  &W^{(2)}_{\psib A} = \frac{\nabla_{\psib}\mathcal{C}_A^-}{\tau}Z^\sigma \nn\\
  &+\Rc\left[-2\sg_A{}^\theta\rho_{\psib}+\sin\theta\sg_A{}^\phi
    \Rc\nabla_{\psib}h_\times-\sg_A{}^\theta\Rc\nabla_{\psib}h_+\right]
  Z^\theta\nn\\
  &+\sin^2\theta\Rc\left[-2\sg_A{}^\phi\rho_{\psib}+\sg_A{}^\theta
    \frac{\Rc}{\sin\theta}\nabla_{\psib}h_\times+\sg_A{}^\phi
    \Rc\nabla_{\psib}h_+\right]Z^\phi\,,\nn\\
  &W^{(2)\theta\theta}=\frac{2}{\tau\Rc}Z^{\sigmab}
  -\frac{\nabla_{\psib}\mathcal{C}_\theta^+}{\tau}Z^\theta\nn\,,\\
  &W^{(2)\theta\phi}=-\frac{\nabla_{\psib}\mathcal{C}_\theta^+}{\tau}Z^\phi\,,
\end{align}
and all the unmentioned components are zero. This means that these
SNFs can now be multiplied by the~$\Rc^{1+n}$ that comes from the LHS
of the first equation in~\eqref{finalfinal1} or the~$\Rc^{3}$ in the
second one and still not have diverging factors. We have now used up
all the freedom we had in choosing gauge, adding constraints and
picking equations and reduction variables, so the set of equations is
finally complete. Equations~\eqref{finalfinal1},~\eqref{finalfinal2}
and~\eqref{finalfinal3} are a system of~$55$ first order differential
equations that contain a few formally singular terms which will take
finite limits on null infinity,~\eqref{formallysing1}
and~\eqref{formallysing3}.

\section{Hyperbolicity}
\label{section:Hyperbolicity}

Let us define a vector~$v^a$ whose entries are each of the~$55$
reduction variables,
\begin{align}
  \mathbf{v} =
  (\Phi_{\psib},\Phi_\psi,\Phi_A,\Phi,...,\rho_{\psib},\rho_\psi,\rho_A,\rho)^T\,.
\end{align}
In order to show hyperbolicity we only need to look at the principal
part so, by definition, we can disregard all the SNFs. Moreover,
although we want to show that the
system~\eqref{finalfinal1},~\eqref{finalfinal2}
and~\eqref{finalfinal3} is hyperbolic, we choose to work with the
system in the second coordinate system~\eqref{2ndCoords}, because it
is simpler and there is a straightforward way to show that
hyperbolicity gets carried over.

\paragraph*{Hyperbolicity in second coordinate system:} We take the
equations before any change of coordinates,
\eqref{EqsRescaled},~\eqref{EqGaugeDriverRescaled}
and~\eqref{EqsRescaledRc} and see that, as the the first coordinate
change leaves the directional derivatives untouched, replacing the
standard radius with the areal radius leaves us with essentially the
same equations. Splitting the null derivatives as combinations of
coordinate basis vectors we can write the principal part of our system
as the following,
\begin{align}\label{hyperbolicityUppercase}
	\mathbf{A}^{\mathring{T}}\p_{\mathring{T}} \mathbf{v}=
        \mathbf{M}^{\mathring{p}}\p_{\mathring{p}}\mathbf{v}+\mathbf{S}\,,
\end{align}
where~$\mathbf{S}$ denotes all the non-principal terms
and~$\mathbf{M}^{\mathring{p}}$ and~$\mathbf{A}^{\mathring{T}}$ can be
written as block diagonal matrices,
\begin{align}
  \mathbf{M}^{\mathring{p}} &=
  \textrm{diag}(P^{\mathring{p}}\,,\dots, P^{\mathring{p}})\,,\nn\\
  \mathbf{A}^{\mathring{T}} &= \textrm{diag}(Q\,,\dots, Q)\,.
\end{align}
Here all entries left blank are zero and~$P^{\mathring{p}}$ and~$Q$
can be written as,
\begin{align}
  P^{\mathring{p}}=\begin{bmatrix}
  -\cpr s^{\mathring{p}} & 0 & \frac{\tau e^\varphi}{2}\sg^{\mathring{p}A} & 0\\
  0 & -\cmr s^{\mathring{p}} & \frac{\tau e^\varphi}{2}\Rc\sg^{\mathring{p}A} & 0\\
  \frac{\cpr}{\tau}\sg_A{}^{\mathring{p}} &
  -\frac{\cmr}{\tau\Rc}\sg_A{}^{\mathring{p}} & \alpha s^{\mathring{p}} & 0\\
  0 & 0 & 0 & -\cmr s^{\mathring{p}}
  \end{bmatrix}\,,
\end{align}
and,
\begin{align}
  Q=\begin{bmatrix}
  1 & 0 & -\frac{\tau e^\varphi}{2\mathring{\tau}}\sg^{AB}\mathring{\mathcal{C}}_B & 0\\
  0 & 1 & -\frac{\tau e^\varphi}{2\mathring{\tau}}\Rc\sg^{AB}\mathring{\mathcal{C}}_B & 0\\
  -\frac{\cpr}{\tau\mathring{\tau}}\mathring{\mathcal{C}}_A &
  \frac{\cmr}{\tau\mathring{\tau}\Rc}\mathring{\mathcal{C}}_A & 1 & 0\\
  0 & 0 & 0 & 1
	\end{bmatrix}\,,
\end{align}
where~$s^{\mathring{p}}=\p_{\mathring{R}}{}^{\mathring{p}}$
and~$\alpha:=-(\p_T\Rc +\frac{\Rc
  e^\varphi\cpr}{2\tau}Z^{\sigmab})$. We want to show that our
equations are symmetric hyperbolic and that requires finding a
matrix~$\mathbf{D}$ such that~$\mathbf{D}\mathbf{M}^{\mathring{p}}$
and~$\mathbf{D}\mathbf{A}^{\mathring{T}}$ are symmetric
and~$\mathbf{D}\mathbf{A}^{\mathring{T}}$ is positive definite. We try
the simplest possible ansatz, a diagonal matrix and we find that a
matrix of the form,
\begin{align}
  \mathbf{D} = \textrm{diag}(\tilde{P}\,,\dots, \tilde{P})\,,
\end{align}
and find that,
\begin{align}
  \tilde{P} = \begin{bmatrix}
    \frac{2e^{-\varphi}}{\tau^2}\cpr& & & \\
    &-\frac{2e^{-\varphi}}{\tau^2\Rc^2}\cmr& &\\
    & & \sg^{AB} &\\
    & & &1
  \end{bmatrix}\,,
\end{align}
fulfills these criteria. To see this, we show the
products~$\tilde{P}P^{\mathring{p}}$ and~$\tilde{P}A^{\mathring{T}}$,
\begin{align}\label{HM1}
  &\tilde{P}P^{\mathring{p}} = \\
  &\begin{bmatrix}
     -\frac{2e^{-\varphi}}{\tau^2}(\cpr)^2 s^{\mathring{p}} & 0 &
     \frac{\cpr}{\tau}\sg^{\mathring{p}A} &0 \\
     0 & \frac{2e^{-\varphi}}{\tau^2\Rc^2}(\cmr)^2 s^{\mathring{p}} &
     -\frac{\cmr}{\tau\Rc}\sg^{\mathring{p}A} & 0\\
     \frac{\cpr}{\tau}\sg^{B\mathring{p}}
     & -\frac{\cmr}{\tau\Rc}\sg^{B\mathring{p}}&
     \alpha s^{\mathring{p}}\sg^{AB} &0\\
     0& 0& 0&-\cmr s^{\mathring{p}}
   \end{bmatrix},\nn
\end{align}
and,
\begin{align}\label{HM2}
  &\tilde{P}A^{\mathring{T}} = \\
  &\begin{bmatrix}
     \frac{2e^{-\varphi}}{\tau^2}\cpr & 0 &
     -\frac{\cpr}{\tau\mathring{\tau}}\sg^{AB}\mathring{\mathcal{C}}_B &0 \\
     0 & -\frac{2e^{-\varphi}}{\tau^2\Rc^2}\cmr &
     \frac{\cmr}{\tau\mathring{\tau}\Rc}\sg^{AB}\mathring{\mathcal{C}}_B & 0\\
     \frac{\cpr}{\tau\mathring{\tau}}\sg^{AB}\mathring{\mathcal{C}}_B&
     \frac{\cmr}{\tau\mathring{\tau}\Rc}\sg^{AB}\mathring{\mathcal{C}}_B&
     \sg^{AB} &0\\
     0& 0& 0&1
   \end{bmatrix},\nn
\end{align}
and check that these matrices are indeed symmetric. Furthermore, using
computer algebra is is straightforward to compute the eigenvalues of
the matrix~\eqref{HM2} and see that it is positive-definite. We then
conclude that our system of equations in the
coordinates~$(\mathring{T},\Rc,\mathring{\theta}^A)$ is symmetric
hyperbolic.

\paragraph*{Coordinate change:} Changing the coordinates
of~\eqref{hyperbolicityUppercase} to the compactified
ones~\eqref{3rdCoords} we get,
\begin{align}
  &\mathbf{A}^{\mathring{T}}J_{\mathring{T}}{}^\alpha\p_\alpha \mathbf{v} =
  \mathbf{M}^{\mathring{p}}J_{\mathring{p}}{}^\alpha\p_\alpha \mathbf{v}
  + \mathbf{S}\\
  \Rightarrow& \mathbf{X}\p_t \mathbf{v} =
  (\mathbf{M}^{\mathring{p}}J_{\mathring{p}}{}^p
  -\mathbf{A}^{\mathring{T}}J_{\mathring{T}}{}^p)\p_p \mathbf{v}
  + \mathbf{S}\nn\,,
\end{align}
where~$\mathbf{X}:=\mathbf{A}^{\mathring{T}}J_{\mathring{T}}{}^t
-\mathbf{M}^{\mathring{p}}J_{\mathring{p}}{}^t$
and~$J_{\mathring{\alpha}}{}^\alpha$ are the components of the
Jacobian associated to the
change~$(\mathring{T},\Rc,\mathring{\theta}^A)\rightarrow
(t,r,\bar{\theta}^A)$. We can then write that,
\begin{align}\label{hyperbolicityLowercase}
  \p_t\mathbf{v}= \bar{\mathbf{M}}^p\p_p \mathbf{v}+\bar{\mathbf{S}}\,,
\end{align}
where~$\bar{\mathbf{M}}^p$ is given by,
\begin{align}
  \bar{\mathbf{M}}^p =
  \mathbf{X}^{-1}(\mathbf{M}^{\mathring{p}}J_{\mathring{p}}{}^p
  -\mathbf{A}^{\mathring{T}}J_{\mathring{T}}{}^p)\,.
\end{align}
It can be seen that the Jacobian of the change of coordinates and its
inverse are,
\begin{align}
  J=\begin{bmatrix}
  1&0& 0 \\
  H'\Rc'&\Rc'&0 \\
  \p_{\mathring{\theta}^A}H&0 & 1
  \end{bmatrix},\,\quad J^{-1}=\begin{bmatrix}
  1&0 & 0 \\
  -H'&\frac{1}{\Rc'}&0 \\
  -\p_{\mathring{\theta}^A}H&0 & 1
  \end{bmatrix},
\end{align}
respectively, so that,
\begin{align}
  \bar{\mathbf{M}}^p = \mathbf{X}^{-1}\mathbf{M}^{\mathring{p}}J_{\mathring{p}}{}^p\,.
\end{align}
We claim that the matrix~$\mathbf{D}\mathbf{X}$ diagonalizes the new
matrix~$\bar{\mathbf{M}}^p$ and to show that we only need to prove
that~$\mathbf{D}\mathbf{X}\bar{\mathbf{M}}^i$ is symmetric and
that~$\mathbf{D}\mathbf{X}$ is positive definite. It is easy to see
that~$\mathbf{D}\mathbf{X}\bar{\mathbf{M}}^p$ is indeed symmetric, so
let us analyze the second condition. We have that,
\begin{align}\label{H(1+X)}
  \mathbf{D}\mathbf{X}
  = \mathbf{D}\left[\mathbf{A}^{\mathring{T}} + \mathbf{M}^{\Rc}H'
    +\mathbf{M}^{\mathring{\theta}^A}(\p_{\mathring{\theta}^A}H)\right]\,.
\end{align}
Once again using computer algebra to compute the eigenvalues of this
matrix we can easily see that they are all positive. Therefore we
conclude that our system in radially compactified
coordinates,~\eqref{finalfinal1},~\eqref{finalfinal2}
and~\eqref{finalfinal3} is symmetric hyperbolic.

\section{Conclusions}
\label{section:conclusions}

Continuing our research program on the inclusion of null-infinity in
the computational domain, we studied the hyperboloidal
compactification of the dual-frame generalized harmonic gauge
formulation. Our aim was to obtain the most regular form of the
equations of motion with this formulation on the compactified
domain. We have given a procedure which, upon careful choice of gauge
and constraint addition, allows us to turn the Einstein field
equations in generalized harmonic gauge into a set of~$11$ second
order differential equations, $10$ of which fall into the categories
of \textit{good} and \textit{ugly}, whose asymptotic solutions behave
respectively like those of the wave equation or decay faster. The
eleventh is a wave equation satisfied by a gauge driver, an auxiliary
function that is used to sway the asymptotics of one of the metric
functions without spoiling hyperbolicity. As shown
in~\cite{DuaFenGasHil21}, polyhomogeneous expansions of ugly fields
may have logarithmically divergent terms in subleading orders
in~$R^{-1}$. Our method allows us to make sure that those terms are
not generated below a specified order. Therefore, this approach
effectively provides a way to help regularize the equations at null
infinity. As the character of the metric functions is determined
solely by their asymptotics and that of their first derivatives, the
choice of gauge and constraint addition that allows for this to happen
is not unique.

Treating the first derivatives of the original variables as evolved
variables in their own right we transformed the system of~$11$ second
order wave equations into a set of first order equations. We then made
use of the torsion free conditions to find additional equations for
the remaining variables and regarded the definitions of the reduction
variables as reduction constraints and transformed to a radially
compactified hyperboloidal coordinate system. Formally singular terms
were identified in most of the equations and almost all of them were
then canceled through another constraint addition at subleading order,
leaving the final set of equations with only two kinds of simple
formally singular terms that can be dealt with easily using
L'H\^opital's rule. What is more, one of these two types of singular
terms serve the specific purpose of suppressing unphysical radiation
fields that are otherwise present when using harmonic-like gauges. We
concluded by showing that the final system is symmetric
hyperbolic. Despite the presence of formally singular terms, one
strength of the approach from the numerical perspective is that it is
very close to standard formulations used in numerical relativity. We
have thus opened the possibility to use well-established numerical
techniques to treat to the strong field region (including for instance
compact binaries) in conjunction with our proposed setup for
compactification.

In this paper we have focused exclusively on the evolution problem and
ignored completely the question of finding suitable constraint solved
initial data. Nonetheless we note that the constraint equations on
hyperboloidal slices have been studied, see for
example~\cite{And02,Par08,BucPfeBar09,ShiAnsPan13}, and we expect that
such data can be constructed numerically and evolved using the
formulation put forward. Another shortcoming of the derivations here
is that we have viewed the field equations primarily as a set of
partial differential equations rather than
geometrically. Aesthetically it is therefore appealing to give a more
geometric version of the formulation, but that is again work for the
future.

\acknowledgments

The Authors wish to thank Alex Va\~{n}\'o-Vi\~{n}uales, Christian
Peterson B\'orquez and Shalabh Gautam for helpful discussions. MD
acknowledges support from FCT (Portugal) program PD/BD/135511/2018, DH
acknowledges support from the FCT (Portugal) IF Program IF/00577/2015,
PTDC/MAT- APL/30043/2017. JF acknowledges support from FCT (Portugal)
programs PTDC/MAT-APL/30043/2017, UIDB/00099/2020. EG acknowledges
support from FCT (Portugal) investigator grant 2020.03845.CEECIND.

\normalem
\bibliography{RegHypDF}

\end{document}